\documentclass[12pt]{article}
\usepackage[latin1]{inputenc}
\usepackage[T1]{fontenc}
\usepackage{natbib}
\bibpunct[: ]{(}{)}{;}{a}{}{,} 
\usepackage{array}
\usepackage{amsmath}
\usepackage{amssymb}
\usepackage{array}
\usepackage{epsfig}
\usepackage{textcomp}
\usepackage{setspace}
\usepackage{longtable}
\usepackage{lscape}
\usepackage{graphicx}
\usepackage{booktabs}
\usepackage{subcaption}
\usepackage{float}
\usepackage{multirow}
\usepackage{amsthm}
\usepackage[rm,bf,center,tiny]{titlesec}
\usepackage{rotating}
\usepackage{dcolumn} 
\usepackage{url} 
\usepackage{lmodern} 
\usepackage{pdflscape} 
\usepackage{placeins} 

\usepackage[final]{pdfpages}
\usepackage{graphicx}
\usepackage{hyperref}
\usepackage{gensymb}
\usepackage{listings}

\titlelabel{\thetitle. \enspace}
\titlelabel{\thetitle. \enspace}

\renewcommand\thesubsection{\Alph{subsection}}
\titleformat{\subsection}{\centering\normalfont\it}{{\thesubsection}.}{.5em}{}
\titleformat{\paragraph}{\normalfont\it}{{\thesubsubsection}}{-.0em}{\quad}

\renewcommand{\thefootnote}{\fnsymbol{footnote}}

\begin{document}
	
	\begin{titlepage}
		\begin{center}
			
			\vspace*{1.cm}
			
			\textbf{\huge The Effects of Air Pollution on Teenagers' Cognitive Performance: Evidence from School Leaving Examination in Poland}
			
			\vspace{0.3cm}
			
			\large Agata Danuta Ga{{\l}}kiewicz\protect\footnote{Agata Ga{{\l}}kiewicz, University of Potsdam, IAB N\"urnberg, e-mail: agata.galkiewicz@uni-potsdam.de.}
			
			\vspace{\fill}
			
			\textbf{Abstract}
		\end{center}
		
		\singlespacing
		
		\noindent \small Random disturbances such as air pollution may affect cognitive performance, which, particularly in high-stakes settings, may have severe consequences for an individual's productivity and well-being. This paper examines the short-term effects of air pollution on school leaving exam results in Poland. I exploit random variation in air pollution between the days on which exams are held across three consecutive school years. I aim to capture this random variation by including school and time fixed effects. The school-level panel data is drawn from a governmental program where air pollution is continuously measured in the schoolyard. This localized hourly air pollution measure is a unique feature of my study, which increases the precision of the estimated effects. In addition, using distant and aggregated air pollution measures allows me for the comparison of the estimates in space and time. The findings suggest that a one standard deviation increase in the concentration of particulate matter $PM_{2.5}$ and $PM_{10}$  decreases students' exam scores by around 0.07--0.08 standard deviations. The magnitude and significance of these results depend on the location and timing of the air pollution readings, indicating the importance of the localized air pollution measure and the distinction between contemporaneous and lingering effects. Further, air pollution effects gradually increase in line with the quantiles of the exam score distribution, suggesting that high-ability students are more affected by the random disturbances caused by air pollution.
		
		\renewcommand{\arraystretch}{1}
		\renewcommand{\baselinestretch}{1.3}
		
		\vspace{\fill}
		\vspace{\fill}
		
		\vspace{\fill}
		
		\renewcommand{\arraystretch}{.9}
		\renewcommand{\baselinestretch}{.9}

		\renewcommand{\arraystretch}{1}
		\renewcommand{\baselinestretch}{1.3}
		
	\end{titlepage}
	
	\clearpage
	\renewcommand{\thefootnote}{\arabic{footnote}}
	\setcounter{footnote}{0}
	\setcounter{page}{1}
	
	\onehalfspacing
	

\section{Introduction}
\label{Intro}

According to the World Health Organization (WHO), in 2019, as much as 99\% of the world's population was living in places where the recommended air pollution thresholds were exceeded.\footnote{For details see: \url{https://www.who.int/news-room/fact-sheets/detail/ambient-(outdoor)-air-quality-and-health}.} While the negative impacts of low air quality on health and mortality have been widely demonstrated in the literature \citep{alexander2022impact, anderson2020wind, barreca2021long, bishop2023hazed, klauber2024killing, margaryan2021low, pestel2021health}, recent studies show that air pollution can also influence non-health-related outcomes such as cognitive performance \citep{bedi2021particle, palacios2022indoor}, productivity \citep{chang2019effect, lichter2017productivity, neidell2023air} and human capital formation \citep{bharadwaj2017gray}. At the same time, the effects of air pollution on cognition in high-stakes settings are still understudied, with, to the best of my knowledge, just four papers examining this issue \citep{andersen2024air, carneiro2021effects, ebenstein2016long, roth2016contemporaneous}. Understanding these impacts is crucial as performance under high-stakes conditions may have long-lasting implications for the individuals involved and, from society's point of view, lead to inefficient allocation of resources or productivity losses.

This paper evaluates the short-term effects of air pollution on the cognitive performance of 14--15-year-olds sitting their school leaving examinations in Poland. It uses school-level panel data from 788 primary schools collected over three consecutive school years between 2021 and 2024. For each school, the average exam results in three obligatory subjects (the national language, math, a modern foreign language) is observed. Air pollution readings are drawn from two data sources. First, a \textit{localized} air pollution measure is obtained from measuring devices installed directly in the schoolyard as part of a government-run project. Second, as in most previous studies, air pollution is captured by public air pollution measuring stations located across the country. Both types of device measure two standard air quality indicators -- concentration of particulate matter $PM_{2.5}$ and $PM_{10}$ -- in a continuous manner. To address potential endogeneity of air pollution, I utilize the panel structure of the data by incorporating school and time fixed effects. This aims to capture plausibly random variation in air quality that is independent of schools' fixed characteristics and air pollution time trends. I also include weather-related and time-varying school-related control variables.

The results of my research can be summarized under four key findings. First, I find robust, significant, and negative effects of air pollution on short-term cognitive performance in high-stakes settings. Specifically, the increase in $PM_{2.5}$ and $PM_{10}$ by 10 units ($\mu g/m^{3}$)  reduces students' exam scores by respectively around 0.14 and 0.11 of a standard deviation.\footnote{This result can be interpreted as a 0.07--0.08 of a standard deviation decrease in the exam score for each standard deviation increase in $PM_{2.5}$ or $PM_{10}$. As the exams are graded on a scale of 0--100 points, this decline is equal to a decrease of around 0.1 points.}
Likewise, on days when recommended WHO threshold for concentration of particulate matter $PM_{2.5}$ is exceeded, students' exam scores are 0.20 of a standard deviation lower. Second, the detrimental effects of air pollution are stronger at the higher end of the exam score distribution, suggesting that high-ability students are more strongly impacted by air pollution than their low-ability peers. 
Third, only the localized air pollution measure provides significant estimates of the effects. When air quality is recorded by public air pollution measuring stations, located a few kilometers away from the schools under study, air pollution effects are insignificant and close to zero. This finding indicates the importance of using a precise air pollution measure. 
Fourth, adjusting the time window in which air pollution is measured provides evidence that lingering effects of air pollution, measured several hours before the exam takes place, are as important as the contemporaneous effects experienced at the time of the cognitively demanding activity.

My paper contributes to the literature in several crucial ways. 
First, it supplements a broad body of work on the effects of air pollution on cognitive performance \citep{duque2022coal, gilraine2024jue, heissel2022does}. Specifically, it looks at the short-term effects of air pollution materializing in high-stakes settings, an area that is still understudied \citep{andersen2024air, carneiro2021effects, ebenstein2016long, roth2016contemporaneous}. In so doing, my study also introduces age as a potentially important factor in understanding susceptibility to air pollution. Previous studies focused exclusively on 18--19-year-olds sitting their upper secondary school leaving exams \citep{andersen2024air} or university entrance exams \citep{carneiro2021effects, ebenstein2016long, roth2016contemporaneous}. The individuals in my study are 14--15 years old and due to their younger age, they might be more prone to the harmful effects of air pollution \citep{aithal2023air}. Generally, children are believed to breathe through their mouths more than adults \citep{bateson2008children}, they have less efficient nasal filtering \citep{bennett2008nasal} and their immune system is still developing \citep{bearer1995children, schwartz2004air}. They also spend more time outdoors and are more likely to engage in physical activity \citep{wiley1991study}. If any of these factors were also to prevail among teenagers, at least to some extent, this may lead to a higher absorption of toxic molecules in the airways and lungs, increasing exposure to damage.

Second, my study contributes to the still unresolved discussion on whether low- or high-ability individuals are more prone to the negative effects of air pollution. \cite{andersen2024air} and \cite{zivin2020unintended} found that these effects are significantly stronger at the higher end of the test score distribution. \cite{roth2016contemporaneous} also showed that STEM students are more affected than non-STEM students. In contrast, \cite{carneiro2021effects} and \cite{ebenstein2016long} found that the effects of air pollution are larger among the poorer performers. Since I find a positive association between the magnitude of air pollution effects and the exam score distribution, my results support the hypothesis that more capable individuals experience stronger effects of air pollution on their cognitive performance.

Third, to the best of my knowledge, this study is the first to use hourly air pollution measures directly at the location where the potentially affected individuals are undertaking the cognitively demanding activity. Previous studies used either readings from public air pollution measuring stations located at least a few kilometers away \citep{carneiro2021effects, ebenstein2016long} or used modeling systems to predict the localized air pollution concentration \citep{andersen2024air}. At the same time, the only study using locally accurate measurements is based on one-time air pollution readings, which are highly susceptible to measurement error \citep{roth2016contemporaneous}. The precision of my measure, both in terms of location and time, ensures the reliability of the estimates by minimizing measurement error in the independent variable and consequently reducing attenuation bias, which may have been present in other studies. Further, the hourly data allow us to disentangle the contemporaneous effects and lingering effects of air pollution measured a few hours before performing a cognitively demanding activity.

Overall, my findings provide new evidence that exposure to air pollution causes a deterioration in short-term cognitive performance and the consequences of this effect are not evenly distributed across the society. Therefore, policymakers should take into account that policies aimed at reducing air pollution can also have other positive impacts beyond health-related benefits such as lower mortality or respiratory disease morbidity. Improving air quality may also improve individual productivity and enhance human capital formation. At the aggregated level, it might lead to more efficient allocation of resources and improve the productivity of the overall economy.

The rest of the paper is organized as follows. 
Section II provides background information on the Polish education system, the air pollution typically experienced in Poland, and the Edukacyjna Sie\'c Antysmogowa (ESA) project which involved the installation of air pollution measuring devices directly in the schoolyard of the participating institutions.
In Section III, data and empirical specification are described. 
Section IV presents the empirical results, while Section V benchmarks these results against the findings in the current literature. 
Section VI discusses the sensitivity of air pollution effects depending on the location of the measuring sources and the time window of measurement. 
Section VII looks at the estimated effects from a broader perspective and Section VIII concludes.


\section{Background}
\label{Background}

\subsection{Polish Education System}
\label{Exams}

The structure of the education system in Poland is presented in Figure \ref{school_system}. Obligatory education starts at the age of seven and covers eight years of primary school. The catchment area for each primary school is defined by the law. Children living in this catchment area are guaranteed a place in that school and as such, there is no competition with respect to admission. It is possible to send a child to a school other than the assigned one, but this requires extra effort on the part of the parents including the completion of additional documentation, and is only feasible if the chosen school still has places free. Education in primary school ends with the mandatory school leaving examination.

The next (secondary) educational stage, for pupils aged 15 to 19, is also mandatory. However, young people have free choice as to the type of secondary school they attend.\footnote{Education in Poland is mandatory until the age of 18. Compulsory schooling starts at the beginning of the school year in the calendar year in which a child turns 7. It continues until the end of primary school, but no later than a child's 18th birthday.} The three types of secondary school are: general, technical, and vocational. General secondary school education can be described as academic-track, since it offers students broad theoretical knowledge across different subjects and prepares them to enter higher education. In technical and vocational schools, students obtain knowledge on a more limited number of subjects, although this is combined with the acquisition of practical occupational skills. Alumni of these types of school can enter the labor market immediately. Irrespective of the school type, secondary school students can take the \textit{matura} exam, which is a requirement for entering higher education at university or any other academic institution.

This study focuses on the school leaving examination sat at the end of primary school. These exams are particularly important, as the results play a role in the secondary school admission process. Unlike primary school, secondary school spots are assigned based on the final score obtained by the students and not based on their place of residence. Although the final score is also influenced by school grades as well as additional activities such as volunteering, the results of the primary school leaving exams make up 50\% of the score. Each secondary school defines its own admission thresholds depending on, inter alia, the number of places available and the faced demand. General schools have the highest admission thresholds and vocational schools the lowest. Further, these thresholds also vary widely between schools of the same type. Schools which are viewed as better-performing set higher thresholds, while lower-performing schools opt for lower thresholds. Students are only admitted to their chosen school if their final overall score reaches or exceeds the school's threshold. All in all, this specific setting means primary school leaving exams in Poland can be considered high-stakes examinations that influence future educational paths and career outcomes.\footnote{This conclusion is valid despite the fact that when submitting applications, pupils can only provide an approximation of their final overall score, as they do not receive their exam results until after the recruitment process is completed.}

Primary school leaving exams are taken in three obligatory subjects: the national language (Polish), math, and a modern foreign language (in most cases English).\footnote{In 2022, 97.60\% of all those sitting exams chose English as their modern foreign language, 2.09\% opted for German, and less than 1.00\% for other languages such as Russian, French, Spanish, or Italian.} The exams are organized on three consecutive days in May. They always start at 9.00 a.m. and, depending on the subject, last between 90 and 120 minutes.\footnote{Children with recognized learning difficulties receive additional time to complete the examination of up to 1 hour.} The order of the exams is fixed; children start with Polish on Tuesday, followed by math on Wednesday, and their modern foreign language on Thursday. Although the exams are held at the schools themselves, the examination is standardized and is conducted nationwide, in other words children in all primary schools across the country take the same tests at the same time.

Each subject-specific exam is graded individually and independently of the others pupils sitting the exam (the exams are not graded on the curve). The final exam result is expressed on a scale of 0 to 100 points. Pupils cannot fail the exams in the sense that there is no minimum threshold defining a pass grade. In case of illness or other circumstances preventing a pupil from participating in the examination, there is the opportunity to take the exams approximately one month later in June. Should the pupil be unable to sit their exams on the agreed alternative dates, they have to repeat the grade and sit their examinations in the next school year. Since sitting the exams on the alternative dates in June or in the following school year tends to be the exception, I focus only on the main examination days in May.\footnote{In 2022, 97.8\% of all 8th grade students participated in the school leaving examination in May (97.1\%) or June (0.7\%). The remaining 2.2\% are students who had to repeat the grade and sit the exams the following year, but also those who were officially exempted from the assessment.} Lastly, the examination cannot be repeated once taken, in other words, the result obtained cannot be improved by retaking the exam at a later point in time.

\subsection{Air Pollution in Poland}
\label{Pollution}

\begin{figure}[hbt!]
	\begin{center}
		
		\caption{Air Pollution in the European Union}
		\vspace{0.2cm} 
		\label{in_europa}
		
		\begin{subfigure}{0.85\textwidth}
			\caption{Air Pollution Measured in $PM_{2.5}$}
			\includegraphics[width=\linewidth]{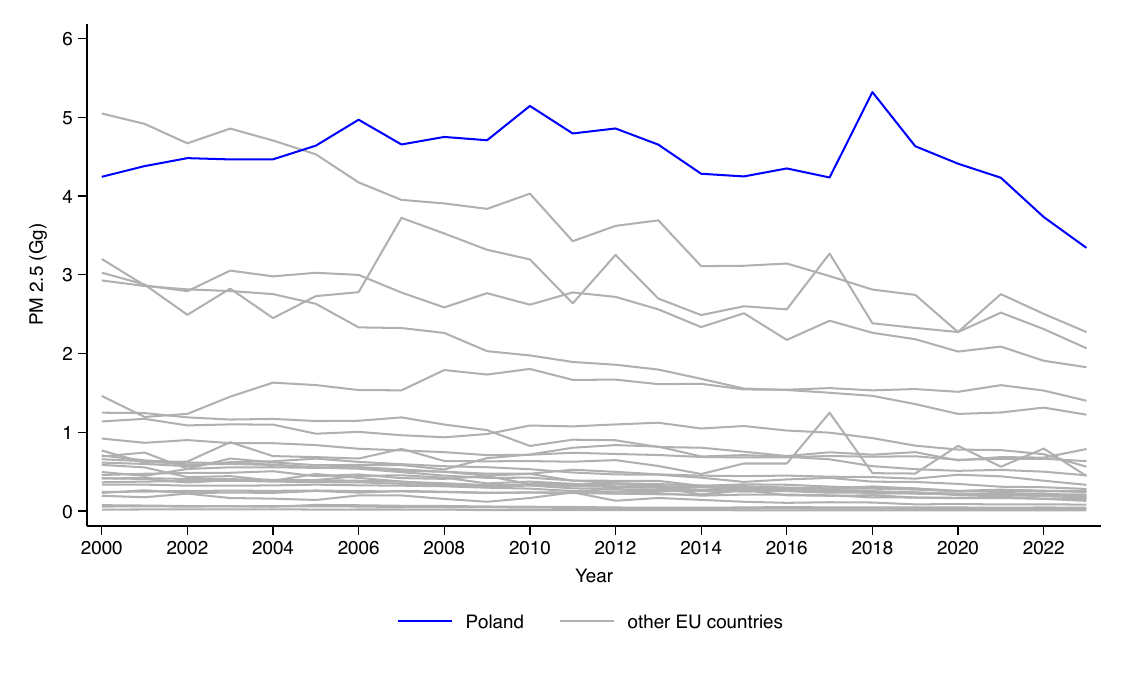}        
		\end{subfigure}
		\hfill		
		
		\begin{subfigure}{0.85\textwidth}
			\caption{Air Pollution Measured in $PM_{10}$}
			\includegraphics[width=\linewidth]{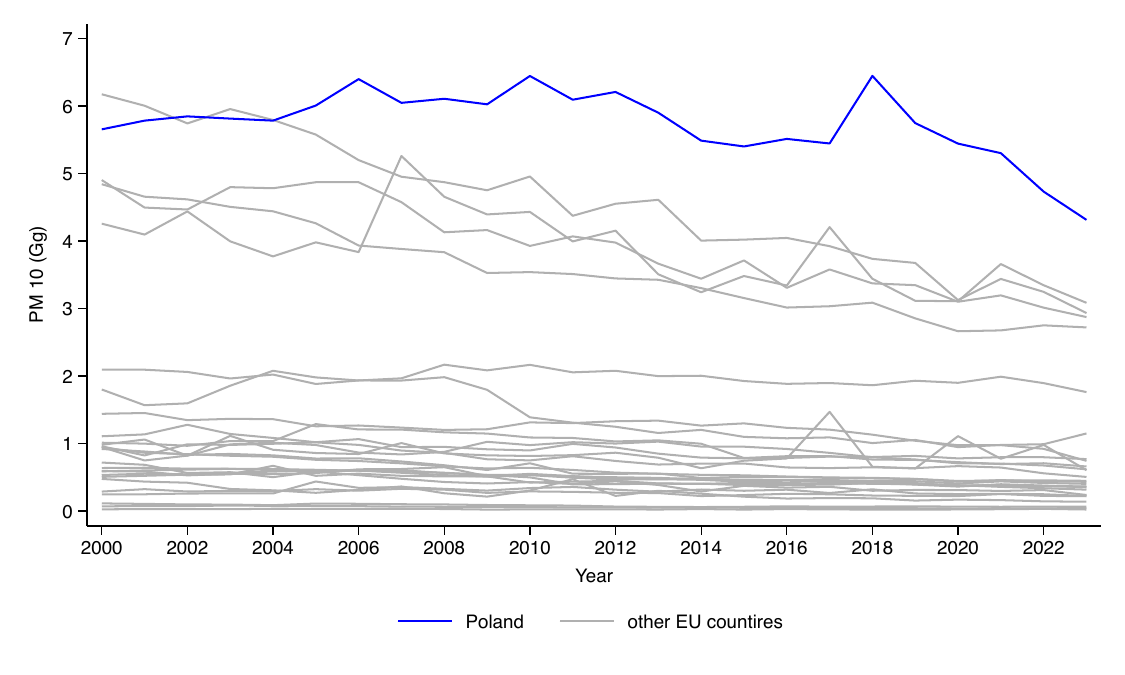}        
		\end{subfigure}
		
	\begin{minipage}{12cm}
		\vspace{0.5cm}
		\footnotesize{
			\textit{Notes}: The figures show the evolution of the amount of air pollution emissions for European Union countries in the time period from 2000 to 2023. The top figure shows (a)	air pollution in $PM_{2.5}$, while the bottom figure (b) shows air pollution in $PM_{10}$. The amount of emissions is expressed in gigagrams (Gg), where 1 Gg is equal to 1,000 tonnes. The blue lines show emissions in Poland, while the gray lines show emissions in other EU countries.}
	\end{minipage}
		
\end{center}
\end{figure}

For long time, Poland has been one of the most polluted countries in Europe \citep{eea_report}. As shown in Figure \ref{in_europa}, the level of $PM_{2.5}$ and $PM_{10}$ emissions has consistently been the highest of all European Union countries throughout the years 2007--2023.\footnote{The comparison is based on the data on emissions of air pollutants reported annually to the European Commission under Directive 2016/2284 of the European Parliament and of the Council on the reduction of national emissions of certain atmospheric pollutants.} According to the European Environment Agency definition, particulate matter is the collective name for fine solid or liquid particles added to the atmosphere by processes occurring at the earth's surface. Particulate matter is not a single molecule, but includes many corpuscles such as dust, smoke, soot, pollen, and soil particles. Of the four types of particulate matter, $PM_{2.5}$ and $PM_{10}$, particles which are respectively 10 or 2.5 $\mu$ or smaller in diameter, are considered inhalable. Because of this characteristic and the potential adverse health effects these particles cause, they are often used as a proxy for overall air quality.

The main sources of high $PM_{2.5}$ and $PM_{10}$ concentration in Poland are ground-level emissions (emissions at a height of up to 40 meters), industry, road transport, and the energy sector \citep{kuchcik2018zanieczyszczenie}. The main reasons Poland is so bad at managing these, rather typical, sources of air pollution compared to other European Union countries is the country's lack of spatial planning and the uncontrolled urban sprawl into suburban areas \citep{iaqm}. For example, the scattered development that characterizes the country makes access to district heating difficult and forces households to meet their heating needs on their own using in-home heating furnaces and local coal-fired boiler rooms. Further, the uncontrolled urban sprawl into rural areas results in an increase in the number of commuters largely using individual cars \citep{ocfp}.

As depicted in Figure \ref{in_europa}, during the years 2018--2023, Poland has seen a gradual improvement in air quality, mostly as a result of changes in legislation. In 2018, the Clean Air Program was introduced with the aim of providing subsidies to those living in single-family buildings for the replacement of outdated and polluting boilers with cleaner heating systems. Further, by 2023 most of the provincial capitals had passed regulations aimed at phasing out high-emission heating sources. Lastly, Poland has been undergoing a slow but steady transition to renewable energy sources \citep{brodny2024empirical}. Despite all these measures, however, recorded levels of $PM_{2.5}$ and $PM_{10}$ still do not comply with WHO recommendations regarding the maximum thresholds for the concentration of these pollutants \citep{nazar2021changes, traczyk2020condition}.

At the same time, as in other Central and Eastern European countries, air quality in Poland follows a clear seasonal pattern. High air pollution is typically observed in the fall/winter season, while the values recorded in the spring/summer season are significantly lower. Temperature inversion, lack of sunlight, weaker wind, and dryer air are the main atmospheric and weather conditions responsible for this phenomenon. Moreover, in countries, like Poland, which rely heavily on solid fuels such as coal and wood, higher air pollution is also experienced in winter due to increased heating demand. Figure \ref{poland_over_year}, which presents the distribution of air pollution in Poland for the years 2022--2023, supports these assertions.\footnote{Figure \ref{poland_over_year} used data from all public air pollution measuring stations.} During the spring/summer season, average air pollution measured in $PM_{2.5}$ amounts to around 12.23 $\mu g/m^{3}$ and measured in $PM_{10}$ to 19.76 $\mu g/m^{3}$. In the fall/winter season in contrast, the average value for $PM_{2.5}$ is 18.38 $\mu g/m^{3}$ and for $PM_{10}$ 25.28 $\mu g/m^{3}$.

\subsection{ESA (Edukacyjna Sie\'c Antysmogowa) Project}
\label{ESA}

The ESA project was launched in 2020. The program is being implemented by the National Research Institute (NASK) together with the Polish Smog Alert (PSA) association. The main goal of the initiative is to raise awareness of the importance of clean air and of the impacts that everyday activities have on air quality. Although this admirable mission pertains to the whole of society, the main target population are children and teenagers attending formal educational establishments. Currently, 2,354 schools are participating in the ESA program.\footnote{This information was taken from the program website (\url{https://esa.nask.pl}), and was last checked on May 19, 2025.}

In most cases, participation in the project is initiated by the schools themselves. To do so, they fill out a dedicated form available on the ESA program website. Since the program is government funded, there are no formal requirements that schools have to fulfill in order to qualify. Therefore, the number of schools admitted to the program is restricted only by the ESA's financial constraints. At the same time, neither central nor local government have any direct influence on whether or not schools participate. Lastly, to achieve representativeness, the program coordinators also actively recruit schools whose characteristics are underrepresented in the ESA sample. The aim of this step is to ensure equality in terms of who profits from the program, despite the lack of information on the individual characteristics of the students. The schools invited to participate receive a letter describing the program and its benefits. In both cases, participation in the ESA program is voluntary and free of charge.

The key aspect of the ESA project for this study is the installation of air pollution measuring devices directly in the schoolyard of the participating establishment. These devices continuously measure $PM_{2.5}$ and $PM_{10}$ as well as temperature, pressure, and humidity. Both weather conditions and air quality are measured directly outside the school. Compared to public air pollution measuring stations, which are based on the reference method, these devices are low-cost, but their quality is exceptionally high.\footnote{The air pollution measuring devices are purchased from the specialized manufacturers and have to comply with predetermined requirements. They exhibit higher precision than off-the-shelf devices.} The devices are calibrated in accordance with the reference values and their exact location, including height or direction of the world, is determined scientifically. All devices are continuously monitored by the program coordinators and undergo annual maintenance. Lastly, malfunctioning devices are repaired or replaced, ensuring continuity of measurement.

All air pollution measuring devices are wirelessly connected to the central server. The values collected are transmitted to the server at five-minute intervals where the data are validated and aggregated to create a moving average from the last 60 minutes. This outcome is updated every five minutes and can be accessed both online and on special LED screens installed in the schools. In addition, historical records can be checked and downloaded free of charge on the program's website. This means that information on air pollution is easily available -- pupils, parents, and school staff can always access it.


\section{Data and Empirical Strategy}
\label{Data_and_Strategy}

\subsection{Data}
\label{Data}

The data on the exam results are recorded at the school level, that is for each school in the sample, average exam results for each of the three obligatory subjects are observed.\footnote{Individual-level data are unfortunately not available.} The data is drawn from the Central Examination Board and can be accessed on the Ministry of National Education homepage. This data is publicly available for the school years 2020/21--2023/24.\footnote{The first primary school leaving exams at the end of 8th grade were sat in May 2019 after the implementation of an extensive educational reform. Exam results from school years 2018/19--2019/20 are only available at the municipality level.} Since the Central Examination Board is also the main body in charge of assessing and grading the exams, the risk of misreported results or nonrandom missing information is limited.

The data on air pollution is drawn from two different sources. In the main analysis, air pollution measurements stem from readings provided by the air pollution measuring devices installed in the schoolyard as part of the ESA program. Here, average hourly values as well as daily means are reported. This data is freely accessible on the ESA project website. In the additional analysis, air pollution data from the closest public air pollution measuring station to a given school is utilized. Here, too, the average hourly and daily measures are available. This data is publicly available on the website of the Chief Inspectorate for Environmental Protection. For both datasets, the manipulation of the data by school staff or municipality government members is very unlikely. Air pollution values measured by the ESA devices are automatically transmitted to the central server with no human involvement. Public air pollution measuring stations, on the other hand, are run by the central authority according to strict legal regulations.

The availability of \textit{localized} air pollution measures from the schoolyard is the key innovation of this paper compared to the existing literature, which largely uses distant air pollution measurements obtained from public air pollution measuring stations located at least a few kilometers away from the potentially affected individuals. In addition, the availability of data on hourly air pollution, compared to the widely used daily average, presents a unique advantage. Thanks to this level of detail, the time window in which air quality is measured can be adjusted to estimate both the contemporaneous and lingering effects of air pollution. The two above features make it possible to mitigate the problem of measurement error in the independent variable that, under a classical errors-in-variables (CEV) assumption, would lead to a biased and inconsistent estimation of the effects.

The data on weather conditions is also drawn from two different sources. In the main analysis, information on hourly and daily average temperature, pressure, and humidity is available. These parameters are simultaneously measured by the air pollution measuring devices installed within the ESA program. In the additional analysis, weather parameters recorded by public meteorological stations run by the Institute of Meteorology and Water Management are taken into account. Here, information on daily average temperature, precipitation, and cloud cover are available. For each school, readings from the closest meteorological stations are utilized.

Lastly, data on school characteristics comes from the Ministry of National Education. In each year, information on the total number of pupils, the number of 8th grade students (including the share of girls in each case), and the number of exam takers is available for each school.\footnote{The number of 8th grade students may differ from the number of exam takers for three main reasons. First, if a student is unable to take the exam on the main or alternative dates. Second, if a student is certified as needing special education and consequently, exempt from the examination. Third, if a student was a finalist in a subject-specific competition at the provincial and supra-provincial level, and is consequently exempt from taking the exam in this particular subject.} In addition, information on whether a school is publicly funded, whether it is part of a bigger educational establishment (primary and secondary schools can make up a jointly governed school cluster), whether it is aimed at special needs pupils (e.g., those who are socially maladjusted or have physical or intellectual disabilities), and whether it has boarders is available.\footnote{Given the children's young age (7--15 years old), schools with boarding facilities are not popular. Most boarding schools are also special needs schools or sporting excellence centers.}

Due to the limited time period for which school-level data on the exam results are available, and due to the fact that the ESA program was not launched until 2020, the analysis in this paper is confined to school years 2021/22--2023/24 (exams in May 2022, 2023, and 2024).\footnote{In school year 2020/21, only a very limited numbers of schools were involved in the ESA project. Therefore, exams taken in 2021 are not included in the analysis.}

\subsection{Sample and Descriptive Statistics}
\label{Sample}

\begin{figure}
\begin{center}
		
		\caption{Distribution of ESA Schools}
		\vspace{0.2cm} 
		\label{location}
		
		\begin{subfigure}{0.85\textwidth}
			\caption{Location of ESA Schools}
			\includegraphics[width=\linewidth]{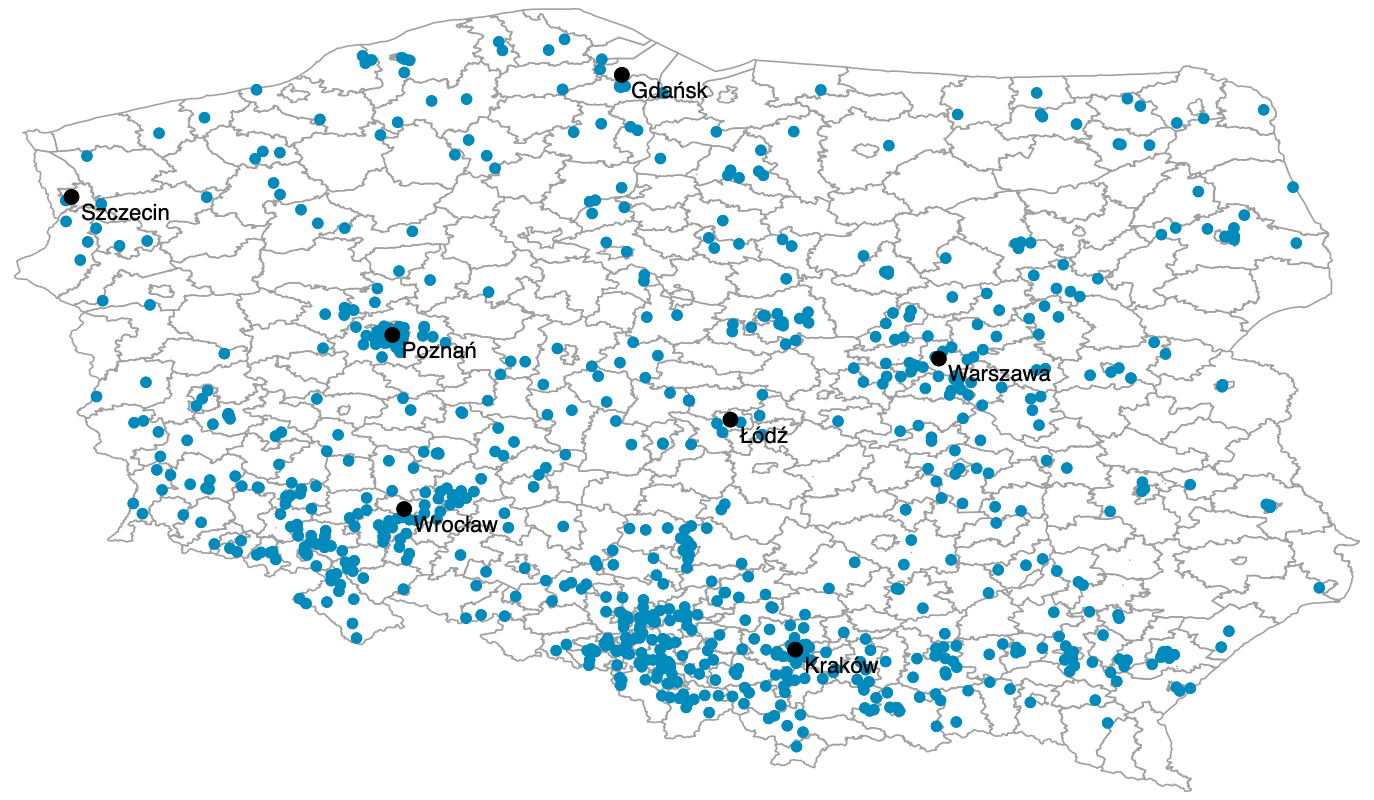}
		\end{subfigure}
		\hfill		
		
		\begin{subfigure}{0.85\textwidth}
			\caption{Population Density}
			\includegraphics[width=\linewidth]{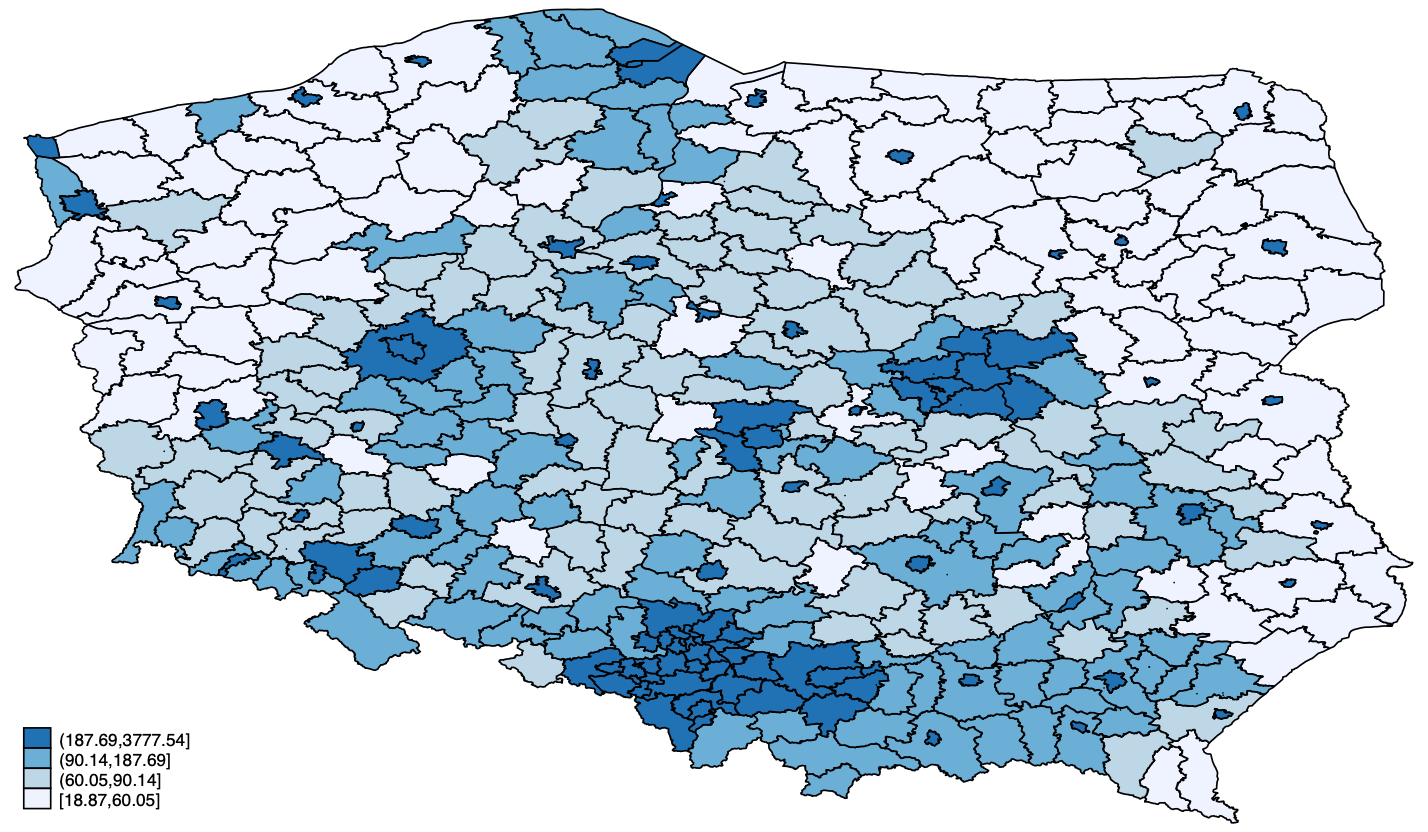}        
		\end{subfigure}
		
	\begin{minipage}{12cm}
		\vspace{0.5cm}
		\footnotesize{
			\textit{Notes}: The upper figure (a) shows the location of primary schools participating in the ESA program and analyzed in this paper. Each school is represented by a blue dot of equal size. Major cities in the country are marked for clarity. The bottom figure (b) shows the population density (number of inhabitants/area in square kilometers) per county in 2021. Darker blue indicates more densely populated areas, while those shaded lighter blue are the less densely populated areas.}
	\end{minipage}
	
\end{center}		
\end{figure}

The analyzed sample comprises 788 of the Polish primary schools participating in the ESA project and for which subject-specific exam results are available.\footnote{Some of the primary schools participating in the ESA project had to be excluded. The most common reasons were that air pollution was not recorded on the days of the exams or the values observed were clearly erroneous. Moreover, for some primary schools, exam results could not be determined. This happens when a school is being restructured and changes its identification number.} The distribution of these schools across the country is presented in Figure \ref{location}. The spatial distribution reflects the population density, which is one of the objectives of the ESA project. As many as 62\% of the municipalities are represented by at least one school in the sample.\footnote{The administrative structure in Poland is as follows: the country is divided into 16 provinces,
380 counties, and 2,477 municipalities.} In these municipalities, the average share of participating schools is around 0.1, with a standard deviation of 0.08.

Table \ref{representativeness} shows the representativeness of the sample by comparing the basic characteristics of the schools participating in the ESA project with the overall population of primary schools in Poland.\footnote{Table \ref{representativeness} is based on the data from school year 2020/21. It is not possible to use the schools' characteristics measured before the launch of the ESA program due to the education system reform that took place between 2019 and 2021 and resulted in untraceable changes in the schools' identification numbers. A subset of very small schools (with up to five pupils sitting the exams) is excluded, as the exam results for these schools are not reported.} The sample of ESA schools is representative, inter alia, in terms of location in urban vs. rural regions and share of female students. The ESA schools are also characterized by slightly better performance on the final exam, are more likely to be located in typically polluted provinces, and have around nine pupils more in the 8th grade. These differences slightly limit the generalizability of my results but do not undermine the internal validity of the study.

Table \ref{sum_stats} presents the summary statistics for the analyzed sample of 788 primary schools, resulting in a total of 4,650 observations. The average exam result in the national language (Polish) was around 62\%, in the modern foreign language (English) around 63\%, and in math around 50\%. Because the distribution of scores in these three different subject-specific exams is slightly different, but the testing regime remains constant, I pool all the observations into one sample and standardize the exam result outcome with a mean of zero and a standard deviation of one \citep{garg2020temperature, rosa2019achievement}.\footnote{Due to the limited number of observations per subject-specific exam and thus limited variation in the air pollution measure, the consideration of each subject-specific exam separately is not credible.} Regarding the independent variable, I measure air quality in five different ways. First, since the ESA devices installed in the schoolyard measure both $PM_{2.5}$ and $PM_{10}$, I consider both of these measures. The mean value of $PM_{2.5}$ and $PM_{10}$ in my sample is 7.74 $\mu g/m^{3}$ and 9.33 $\mu g/m^{3}$, respectively. Next, the Air Quality Index (AQI) is calculated using the $PM_{10}$ values. This is my third air pollution measure.\footnote{The Air Quality Index (AQI) is the U.S. Environmental Protection Agency's air quality measure. It is calculated based on five main pollutant measures, including particulate matter $PM_{2.5}$ and $PM_{10}$.} Lastly, I define two dummy variables, based on $PM_{2.5}$ and $PM_{10}$, which differentiate between days with more and less pollution. The high pollution dummy in the case of $PM_{2.5}$ is defined as 1 if air pollution is equal to or greater than 15 $\mu g/m^{3}$, and 0 otherwise. This threshold is based on WHO air quality guidelines (AQGs) from 2021. The high pollution dummy in the case of $PM_{10}$ is defined as 1 if air pollution is equal to or greater than 20 $\mu g/m^{3}$, and 0 otherwise.\footnote{The WHO air quality guidelines (AQGs) set a threshold value of 45 $\mu g/m^{3}$ for $PM_{10}$. This value is almost never reached in the data, however.} The ESA air pollution measuring devices also provide information on the outside temperature, pressure, and humidity. The average values are indicated in Table \ref{sum_stats}.

\begin{figure}
\begin{center}

	\centering    
	\caption{Distribution of Air Pollution}
	\label{distribution}
	\vspace{0.2cm}
	\centering
	
	\begin{subfigure}{0.48\textwidth}
		\centering
		\caption{Sample Distribution ($PM_{2.5}$)}	
		\includegraphics[width=\textwidth]{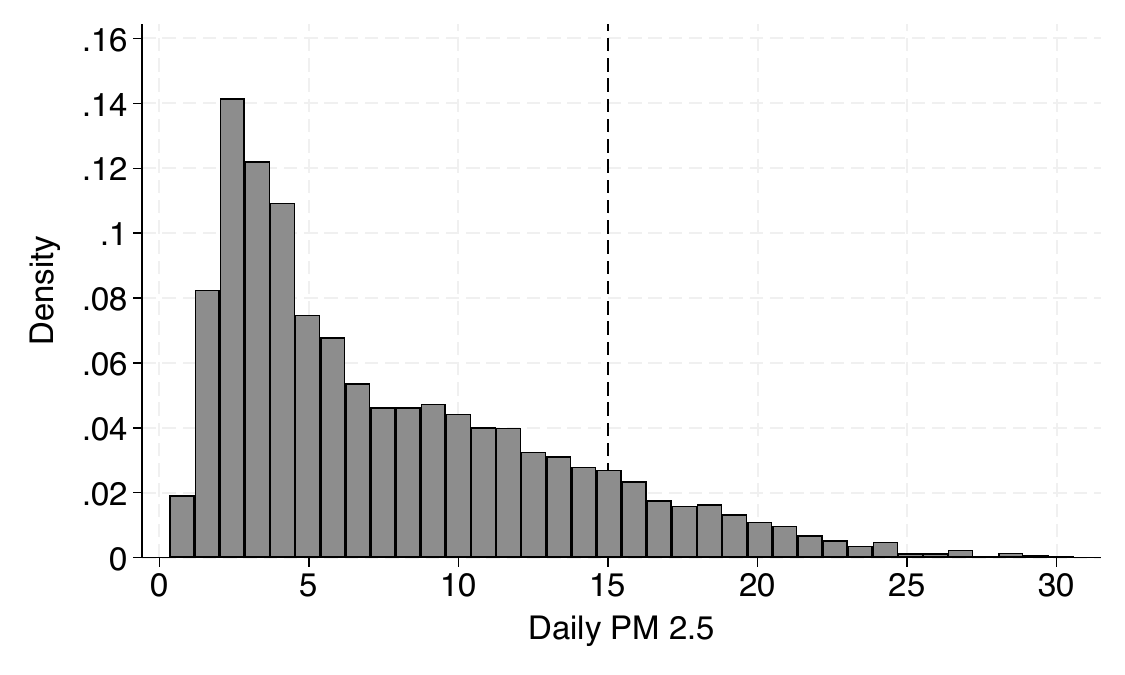}
	\end{subfigure}
	\hspace{0.2cm}
	\begin{subfigure}{0.48\textwidth}
		\centering
		\caption{Sample Distribution ($PM_{10}$)}
		\includegraphics[width=\textwidth]{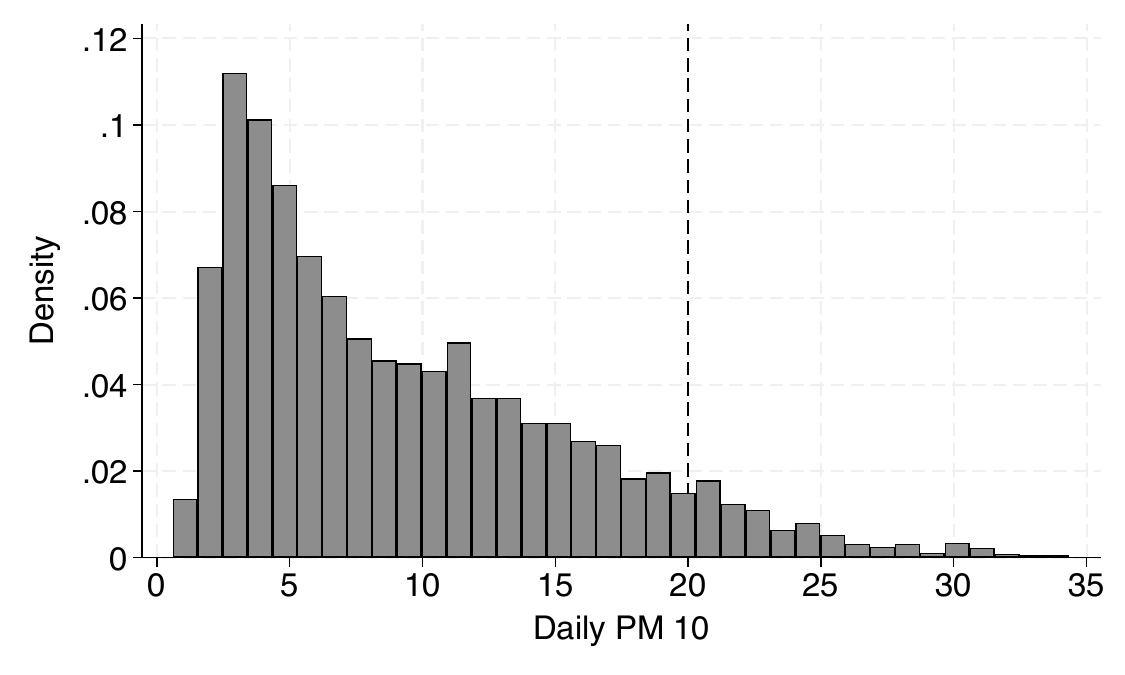}
	\end{subfigure}
	
	\begin{subfigure}{0.48\textwidth}
		\centering
		\caption{Distribution within Schools ($PM_{2.5}$)}
		\includegraphics[width=\textwidth]{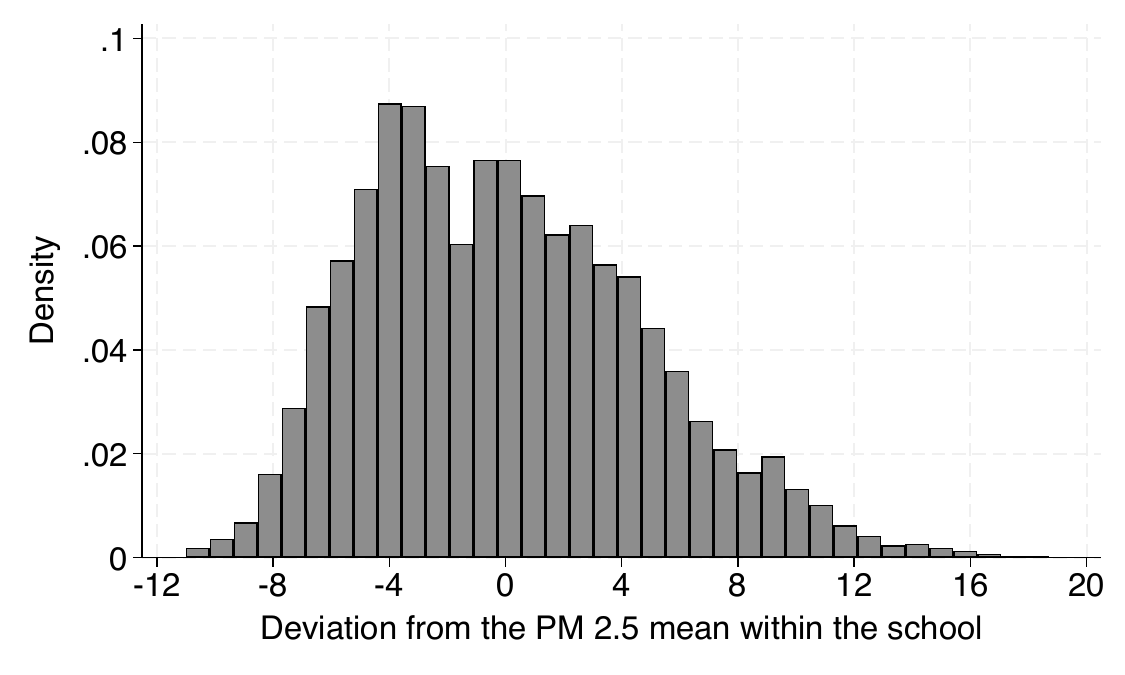}
	\end{subfigure}
	\hspace{0.2cm}
	\begin{subfigure}{0.48\textwidth}
		\centering
		\caption{Distribution within Schools ($PM_{10}$)}
		\includegraphics[width=\textwidth]{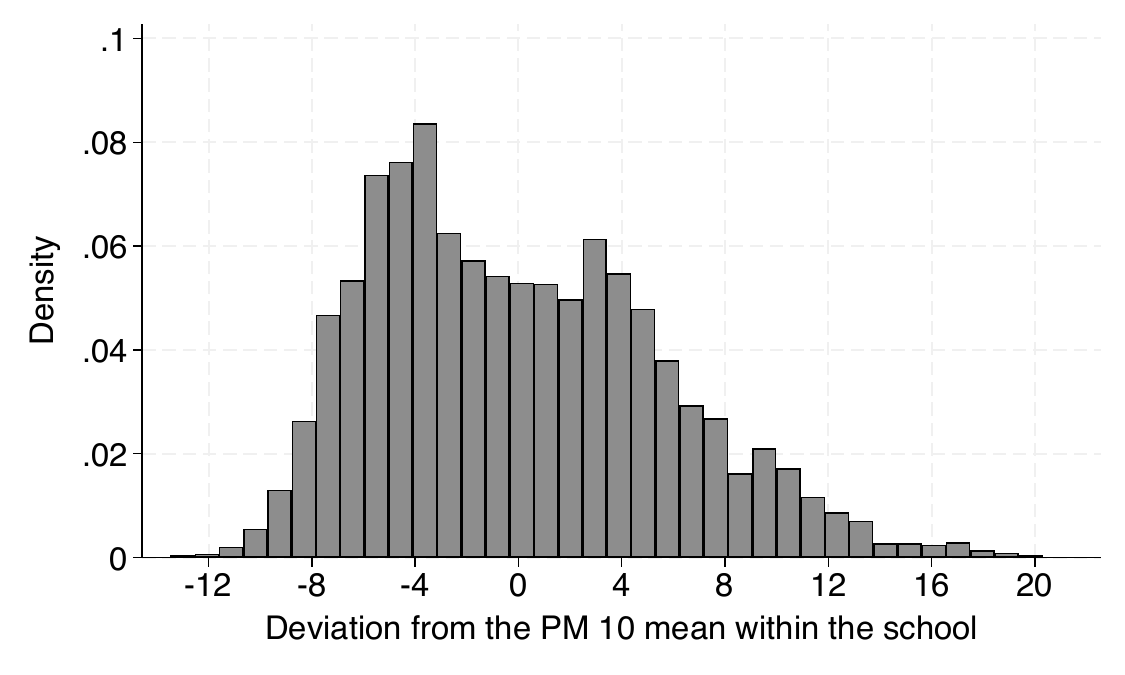}
	\end{subfigure}
	
	\begin{minipage}{12cm}
		\vspace{0.5cm}
		\footnotesize{
			\textit{Notes}: The figures show the distribution of air pollution in the sample. The top figures show the density of daily air pollution measured in $PM_{2.5}$ (a) and in $PM_{10}$ (b). The dashed lines indicate the thresholds used to define binary variables. For $PM_{2.5}$, the threshold is 15 $\mu g/m^{3}$, while for $PM_{10}$, it is 20 $\mu g/m^{3}$. The bottom figures show the density of deviations from the school's air pollution mean, measured in $PM_{2.5}$ (c) and in $PM_{10}$ (d).	The mean air pollution within the school is calculated using all available observations for that school.}
	\end{minipage}
	
\end{center}
\end{figure}

Although the average air pollution values in the sample are not high, the air pollution distribution ensures that there should be enough variation to estimate the causal effects. The top graph in Figure \ref{distribution} shows the density of air pollution in the analyzed sample over the exam days. The bottom graph in Figure \ref{distribution} presents the air pollution distribution within the observed schools over the exam days. A formal test to check for the sufficient amount of variation is also performed and successfully passed.\footnote{The formal test is based on the $R^{2}$ values and, in this case, involves regressing air pollution on other independent variables used in the main specification (including school fixed effects, time fixed as well as weather- and school-related controls) \citep{bedi2021particle, carpenter2008cigarette}. While an $R^{2}$ above the value of 0.9 is perceived as problematic, for $PM_{2.5}$,
the $R^{2}$ value recorded was 0.805 and for $PM_{10}$ 0.789.}
To further visualize the variation used in my model, Figure \ref{example} plots the time series of air pollution (measured as $PM_{2.5}$ and $PM_{10}$) across the examination days for randomly selected schools within predefined clusters. The four clusters were defined based on urbanization status (urban vs. rural area) and typical pollution (polluted vs. clean provinces). Within each cluster, three schools -- one from each of the three quantiles of the school size distribution -- were randomly chosen. As Figure \ref{example} shows, irrespective of the school's location and size, there is sufficient variation in air quality between the different exam days.\footnote{This conclusion also holds if schools are randomly chosen in other ways.}

\subsection{Empirical Specification}
\label{Strategy}

To answer the question on the effects of air pollution on cognitive performance during the primary school leaving examination, the following empirical model is estimated:
\begin{equation}
	Y_{st} = \beta AP_{st}+f(T_{st},H_{st},P_{st})+\pi X_{st}+\psi _{s}+\lambda _{t}+\phi _{t}+\varepsilon _{st}
\end{equation} 
where  $Y_{st}$ is the average subject-specific exam result in school \textit{s} at time \textit{t}, $AP_{st}$ is the measure of air pollution in school \textit{s} at time \textit{t}, and $\beta$ is the coefficient of interest. Depending on the specification, air pollution is measured in five distinct ways: as the concentration of $PM_{2.5}$ or $PM_{10}$ (both expressed in $\mu g/m^3$), as the Air Quality Index (AQI), and as dummy variables based on $PM_{2.5}$ or $PM_{10}$ values and indicating high vs. low air pollution days. $f(T_{st},H_{st},P_{st})$ indicates the function of weather controls, which includes temperature (in \degree C), pressure (in hPa), and humidity (in \%). The main specification controls for linear and quadratic weather controls as well as the interaction terms between them.\footnote{In the additional analysis using air quality measurements taken at the public air pollution measuring stations, the observed weather controls are: temperature (in \degree C), precipitation (in mm), and cloud cover (in octane). Here, linear and quadratic terms, as well as the interaction terms, are also included as control variables.} $X_{st}$ is the vector of time-varying school-related control variables, which includes the number of exam takers, the number of 8th grade students, and the number of pupils (with the last two variables including the share of girls). $\psi _{s}$ indicates school fixed effects, $\lambda _{t}$ year fixed effects, and $\phi _{t}$ day of the week (subject) fixed effects. Lastly, $\varepsilon _{st}$ is an idiosyncratic error term. The standard errors are clustered at the school level.

The structure of the above model was based on the fact that I am unlikely to be able to achieve an unbiased estimate of the effects of air pollution on school leaving exam results in a simple OLS regression. The biggest concern with such an approach is that the air pollution pupils are exposed to may not be exogenous but rather correlated with the error term, creating the problem of omitted variable bias (OVB). For example, children from high socioeconomic status (SES) families may live and therefore attend schools in areas which are less polluted and typically more expensive \citep{banzhaf2008people, currie2014we}. At the same time, high SES parents may be more likely to invest in their children's education, contributing to the pupils' higher exam scores \citep{liu2022socioeconomic}. Further, unobserved avoidance or protective behavior may also confound the OLS estimates \citep{neidell2004air}.

To overcome the above concern regarding air pollution endogeneity, my model includes a rich set of fixed effects (FE), captured thanks to the panel structure of the data. The panel consists of three school years (2021/22--2023/24), and most schools are observed in all these years. Since, in each of these years, three subject-specific exams are held, each school is typically observed nine times.\footnote{Some schools may be observed fewer than nine times. This may happen if the results from a particular subject-specific exam in one of the school years considered are not reported. A second reason for this is that the school only joined the ESA program in May 2022, i.e., after the first considered examination period.} This data structure allows me to control for school fixed effects, year fixed effects, and day of the week (subject) fixed effects. School fixed effects eliminate all time-invariant factors related to the school's characteristics that may influence both the exam results and the amount of air pollution the pupils are exposed to. For example, school fixed effects account for differences in school's geographical location and size, as well as neighborhood demographic characteristics and the spatial development of the local area. The incorporation of year fixed effects allows me to control for national or regional trends that affect all schools in the same year. These might be, for example, changes in educational policies or decreasing trends in air pollution. Lastly, day of the week (subject) fixed effects allows me to account for two further concerns. First, it accounts for the fact that pupils may be experiencing increased physical tiredness or mental exhaustion on the second or third day of the examination due to elevated stress levels or constant tension \citep{lupien2007effects}. Second, it allows me to control for the fact that exams in different subjects might exhibit different levels of difficulty or there may be systematic differences in the types of exam questions asked.

The remaining control variables are included in the model to block the back-door paths that interfere in the estimation of the causal effects of air pollution on cognition. For example, air quality is not only determined by the amount of emissions, but also by the meteorological conditions such as wind direction and speed, temperature, humidity, precipitation, or pressure \citep{czarnecka2008meteorological, elminir2005dependence, jacob2009effect, malek2006meteorological}. At the same time, weather conditions may influence
students' cognitive performance \citep{arceo2024extreme, garg2020temperature}. The existence of these mutual relations suggests that available information on meteorological conditions should be included in the model. On the other hand, the weather-related variables may constitute bad controls if the effect runs in the opposite direction, that is if air pollution influences meteorological conditions. Nevertheless, for my study, which seeks to uncover the short-term effects of air pollution, this is not a major concern. Recent studies show that the effects of air pollution are mostly long-term, meaning that particulate matter can affect the climate, but not the weather on a given day \citep{mayer2000linking, prinn2005effects}.\footnote{According to the World Meteorological Organization (WMO), climate is defined as the weather conditions prevailing in general or over a longer period of time in a given area.}

Further, my model controls for crucial time-varying school characteristics not captured by the school fixed effects. In general, these nonconstant school features relate to the composition and size of the cohort. Although, due to the fixed capacity of the school's premises, there is unlikely to be a high degree of variation in the number of students from year to year, even small changes may matter. It has been established that class size has a significant effect on educational outcomes \citep{chetty2011does, krueger1999experimental}. In addition, the gender composition of a class appears to play a nonnegligible role \citep{hoxby2000peer, whitmore2005resource}. Moreover, the number of students in the exam room can influence the extent to which the room is aired and thus also affect the air quality.

The final concern that applies to my setup is the fact that information on current air pollution is salient and easily available. However, I do not find this problematic for several reasons. First, such information is consistently provided in all schools participating in the ESA project. Therefore, there is no selection with respect to which schools or individuals are able to access this information. Second, my fixed effects (FE) model enables me to control for time-invariant school characteristics such as the sensitivity of the school's staff regarding air pollution issues and consequently the degree of preventive behavior undertaken. Since air quality is known throughout the whole observed time period, the assumption that the degree of preventive behavior is the same in each year is quite likely to be true. Third, preventive behavior on national exam days is limited to medical mask wearing and reduced airing of the room. These actions may, however, also limit the students' oxygen supply, which is believed to negatively affect brain activity and cognitive functions \citep{nasrollahi2024effects, smerdon2022effect}. In addition, wearing a medical mask during the examination was not even mandatory during the COVID-19 pandemic. Lastly, due to severe financial problems faced by most public schools, the main goal of the ESA program is to increase air pollution awareness rather than encourage schools to undertake proactive measures. Overall, we should remain skeptical as to whether there is any heterogeneity in preventive behavior across schools at the time of examination that could induce bias in my estimations.

On the other hand, students' awareness of air quality during such important examination, combined with the extensive knowledge on the consequences of air pollution acquired thanks to the educational component of the ESA program may have psychological effects on the exam takers. Knowing that air pollution is relatively high (low) may impair (boost) children's confidence in their own performance. Therefore, part of the effect found in this study may be driven by the psychological reaction to information provision, and not just the biological reaction of the human body to changing air quality. Disentangling these two mechanisms is, however, beyond the scope of this paper.

Lastly, it is worth noting that although the above empirical model accounts for a rich set of variables that may be related both to the outcome and air pollution, there may still be some unobserved factors contemporaneous to air quality measurements. In response, a rich set of robustness checks and placebo tests is performed and discussed in the following sections.


\section{Results}
\label{Results}	
	
\subsection{Main Results}
\label{Main_Results}

The main estimates are presented in Table \ref{main_results}. The outcome variable is the standardized primary school leaving exam result, while the primary independent variable, shown in the rows of the table, are different air pollution measures. Air pollution is measured in $PM_{2.5}$, $PM_{10}$, as AQI and by dummy variables based on $PM_{2.5}$ and $PM_{10}$ thresholds. The specification in column 1 includes only school and time fixed effects. In column 2, linear weather controls are added, while in column 3, squared weather variables and the interaction terms between them are also included. Estimates in column 4 are drawn from the preferred specification, which also incorporates time-varying school-related control variables. The five coefficients in each specification (column) are taken from a separate regression. The first conclusion from the presented results is that coefficients in subsequent specifications (columns) are relatively similar in magnitude and there are no large differences between them. This makes the assumption that the variation in air pollution is as good as random, after controlling for school and time fixed effects, likely to hold. Therefore, adding further controls does not change the estimated results in a meaningful way.

The effects of air pollution measured in $PM_{2.5}$ (first row) are negative and
statistically significant. The results indicate that a 1 unit ($\mu g/m^{3}$) increase in $PM_{2.5}$ reduces the exam score by 0.014 of a standard deviation. The results for $PM_{10}$ (second row), and AQI (third row) are, as expected, similar, since the AQI is based on the $PM_{10}$ values. Here, the effect of a 1 unit increase in air pollution causes a decline in exam results of 0.011--0.012 of a standard deviation. To put these effects into perspective, I contrast the exam results of the least and the most polluted schools. The average air pollution among the schools in the top 10th percentile of air quality distribution is 2.13 units of $PM_{2.5}$, while for those in the 90th percentile, it is 16.03 units. This 14-unit difference in air pollution would lead to a gap in the exam results of around 0.19 of a standard deviation. When the air pollution measure is based on $PM_{10}$, a gap of 0.18 of a standard deviation is predicted. With respect to the nonlinear effects, the coefficient on the $PM_{2.5}$ dummy variable (row 4) indicates that when air pollution values exceed the WHO air quality threshold (15 $\mu g/m^{3}$), exam results decrease by around 0.20 of a standard deviation. Similarly, if the $PM_{10}$ values are above 20 $\mu g/m^{3}$ (row 5), the exam results are expected to drop by around 0.18 of a standard deviation. All these effects are precisely estimated at the 1\% significance level.

Lastly, to visualize these regression results, Figure \ref{residuals} plots the exam result residuals against the air pollution residuals, both measured in $PM_{2.5}$ and $PM_{10}$. To create these graphs, standardized exam results and air pollution are regressed on school fixed effects, time fixed effects, and the other control variables used in the main specification. The obtained residuals represent the variation in the main independent and dependent variables that is unexplained by these controls. Residual exam results are averaged over 0.1 unit bins of residual air pollution and plotted using lowess bandsmoother. The plots therefore show how plausibly exogenous variation in air pollution is associated with the exam results \citep{ebenstein2016long}. In both cases, with air pollution measured in $PM_{2.5}$ (graph a) and $PM_{10}$ (graph b), the relationship seems to be negative. However, there are a few outlying observations on both ends of the distribution with values oscillating in both directions, meaning the downward trend is not as strong at the edges.

\subsection{Robustness Checks}
\label{Robustness}

To assess the robustness of the above findings, results from slightly adjusted specifications are reported in Table \ref{robustness}. In column 1, standard errors are clustered at the municipality level rather than the school level. This accounts for the possible correlation of errors across schools within the same municipality. In column 2, temperature is measured in bins rather than as a continuous variable. This is done to reduce the sensitivity of the results to the functional form of the temperature control variable. Each of the six bins encompass 2\degree C.\footnote{Temperature bins are defined as follows: $<$12, $<$12,14), $<$14,16), $<$16,18), $<$18,20), $>$=20.}

In columns 3 and 4, sample selection issues are taken into account. In column 3, only public schools are considered. The underlying assumption is that nonpublic schools are financed in a different way and usually have a higher budget. This may in turn lead to more investment in technologically advanced air pollution protection measures than the average public school could not afford. Consequently, the effects of air pollution on exam results may be different in these two samples. In column 4, the weighting procedure based on the iterative proportional fitting algorithm is incorporated. This mechanism performs a stepwise adjustment of sampling weights to achieve known population margins. This procedure accounts for the fact that the analyzed sample of ESA schools does not constitute a fully representative sample of all primary schools in Poland.\footnote{The characteristics of all primary schools in Poland are presented in column 2 of Table \ref{representativeness}.}

Overall, none of the above adjustments to the main specification significantly change the estimated coefficients. Only the weighting procedure reduced the significance level from 1\% to 10\%. Even in this specification, however, the effects are still economically and statistically significant.

Further, in columns 6 and 7, potentially confounding factors related to COVID-19 pandemic are considered. In general, the pandemic should not be an issue for the estimation strategy for several reasons. First, all regulations relating to the education system and the functioning of educational establishments are agreed at the national level and therefore impact all schools equally. Second, examinations in all school years under study took place normally, that is without the restriction of keeping a distance of 1.5 meters between pupils. Third, the differential effects that school closure and distance learning had on subsequent cohorts are controlled for by including year fixed effects.\footnote{In school years 2022/23 and 2023/24, all teaching was in person. In school year 2021/22, distance learning was limited to the time periods 20.12.2021--09.01.2022 and 27.01.2022--20.02.2022.}

Nevertheless, the time-varying influences of the COVID-19 pandemic may still
have had some impact on the obtained estimates. For example, in areas with a higher incidence of COVID-19, students might have been more likely to be placed in quarantine due to the higher probability of them or their classmates being infected.\footnote{Quarantine was imposed on a case-by-case basis and depended on the intensity of contact with the infected person.} This increased likelihood of having to stay at home during the school year might have reduced the effectiveness of learning and the quality of preparation for the school leaving exams. In addition, the regional impact of the pandemic may have influenced the use of different means of transport and contributed to the air quality the pupils were exposed to during the school year. The direction of this influence is however not a priori clear. On the one hand, a higher rate of infection may have led to increased use of private cars and therefore reduced the air quality. On the other hand, a more severe impact of the pandemic might have increased the number of days working from home, thus improving the air quality due to lower traffic volume.

To take into account these potential influences of the COVID-19 pandemic, two different proxies for regional pandemic severity are constructed. The specification in column 5 additionally controls for the average number of COVID-19 cases (per 100,000 inhabitants) in each month of the final school year in the county where a given school is located. The specification in column 6, on the other hand, controls for the rate of positive COVID-19 tests in each month of the final school year in the county where a given school is located. In so doing, both specifications accommodate for the fact that regional differences in the severity of the pandemic may have impacted the observed schools differently in each of the years under study.

Further, adding the COVID-19 pandemic-related control variables did not significantly change the estimated coefficients from the main specification. Therefore, the conclusion that air pollution has negative effects on cognitive performance in a high-stakes setting remains valid.

\begin{figure}
	\begin{center}
		
		\caption{Falsification Test: \\ Effects of Air Pollution on Exam Results}
		\vspace{0.2cm} 
		\label{falsification}
		
		\begin{subfigure}{0.85\textwidth}
			\caption{Air Pollution Measured in $PM_{2.5}$}
			\includegraphics[width=\linewidth]{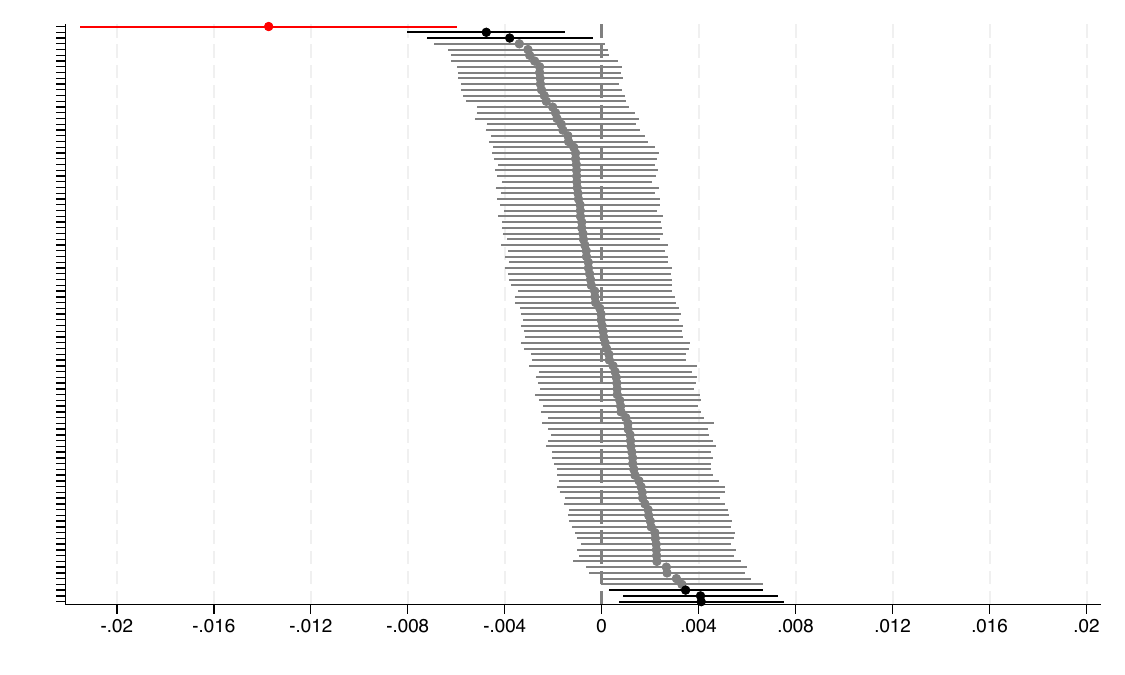}        
		\end{subfigure}
		\hfill		
		
		\begin{subfigure}{0.85\textwidth}
			\caption{Air Pollution Measured in $PM_{10}$}
			\includegraphics[width=\linewidth]{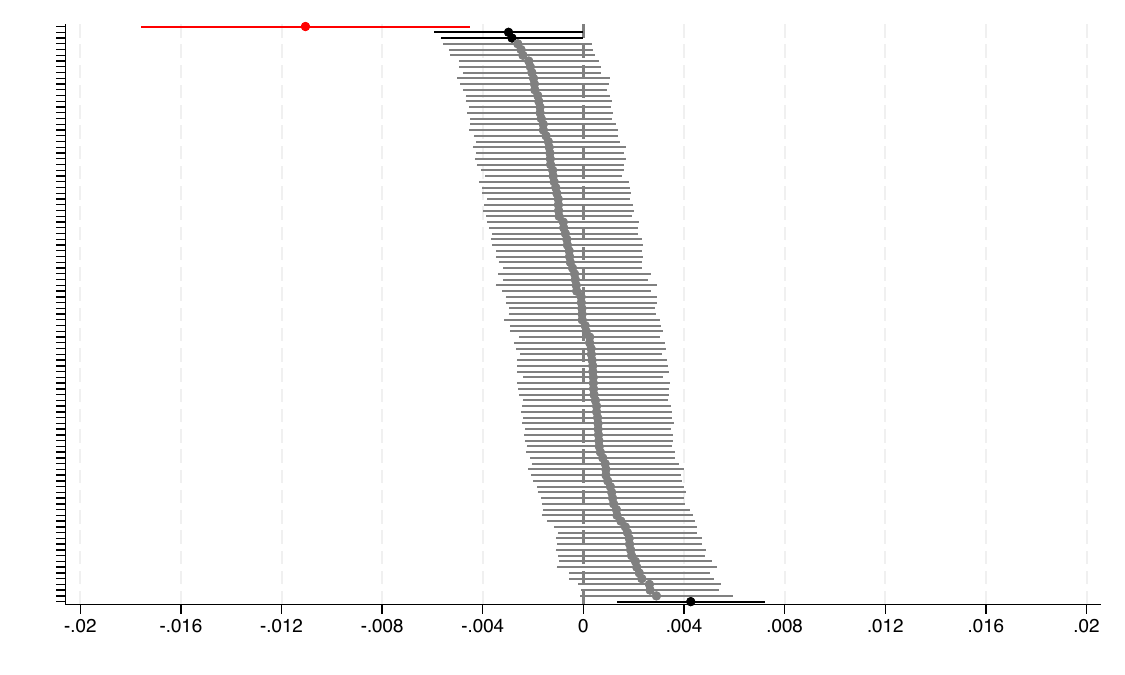}        
		\end{subfigure}
		
	\begin{minipage}{12cm}
		\vspace{0.2cm}
		\footnotesize{
			\textit{Notes}: The figures show the results of the falsification tests. The falsification tests were performed by randomly assigning to each school in the sample air pollution measured in another school on the same exam day. The dark coefficients come from repeating this exercise 100 times. Only black coefficients are significant at the 5\% level. The red coefficients indicate the estimates obtained in the main specification and presented in column 4 of Table \ref{main_results}. Each coefficient is presented together with corresponding 95\% confidence intervals. In the top figure (a), air pollution is measured in $PM_{2.5}$, while in the bottom figure (b), it is measured in $PM_{10}$.}
	\end{minipage}
		
\end{center}
\end{figure}

Lastly, I perform falsification tests in which I assign each school in the sample to the random air pollution value measured in another school, but on the same exam day. This random assignment is performed 100 times and each time the effects of randomly assigned air pollution on the exam results are estimated. The results of this exercise are presented in Figure \ref{falsification}. The upper red coefficients are drawn from the main specification with the correctly assigned air pollution measures.\footnote{These results correspond to those presented in column 4 of Table \ref{main_results}.} The remaining coefficients, ordered by the magnitude, are the outcomes of the random assignment. Only black coefficients are significant at the 5\% level. When air pollution is measured in $PM_{2.5}$, 5\% of the time the falsification test results in significant effects. For air pollution measured in $PM_{10}$, the share of significant coefficients amounts to 3\%. Further, the coefficients from these random assignments are always smaller in magnitude than those from the correct matches. Overall, the outcomes of these falsification tests indicate that the estimated effects of air pollution on exam scores are unlikely to be the result of a simple coincidence.

\subsection{Effect Heterogeneity}
\label{Heterogeneity}

I also perform quantile regression to ascertain whether schools with lower and higher exam results experience the effects of air pollution differently or whether this impact is relatively uniform across the whole population. This allows me to establish whether there are any unwanted social inequalities in terms of air pollution impacts. The existence of such differences would add another layer to the complexity of the air pollution problem.

To date, the evidence presented by the current literature is inconclusive in this regard. On the one hand, \cite{andersen2024air} and \cite{zivin2020unintended} indicate that individuals at the higher end of the score distribution seem to be more strongly affected by air pollution. These conclusions are based on secondary school graduates in Denmark and China. Both results are in line with \cite{roth2016contemporaneous}, who shows that STEM students are more
affected than non-STEM students when looking at the population of university
entrants in London, UK. 
On the other hand, \cite{carneiro2021effects} and \cite{ebenstein2016long} find more harmful effects of air pollution among lower-performing individuals. The first paper studies the population of university candidates in Brazil, while the second looks at university entrants in Israel.

\begin{figure}
	\begin{center}
		
		\caption{Quantile Regressions: \\ Effects of Air Pollution on Exam Results}
		\vspace{0.2cm} 
		\label{quantile_graph}
		
		\begin{subfigure}{0.85\textwidth}
			\caption{Air Pollution Measured in $PM_{2.5}$}
			\includegraphics[width=\linewidth]{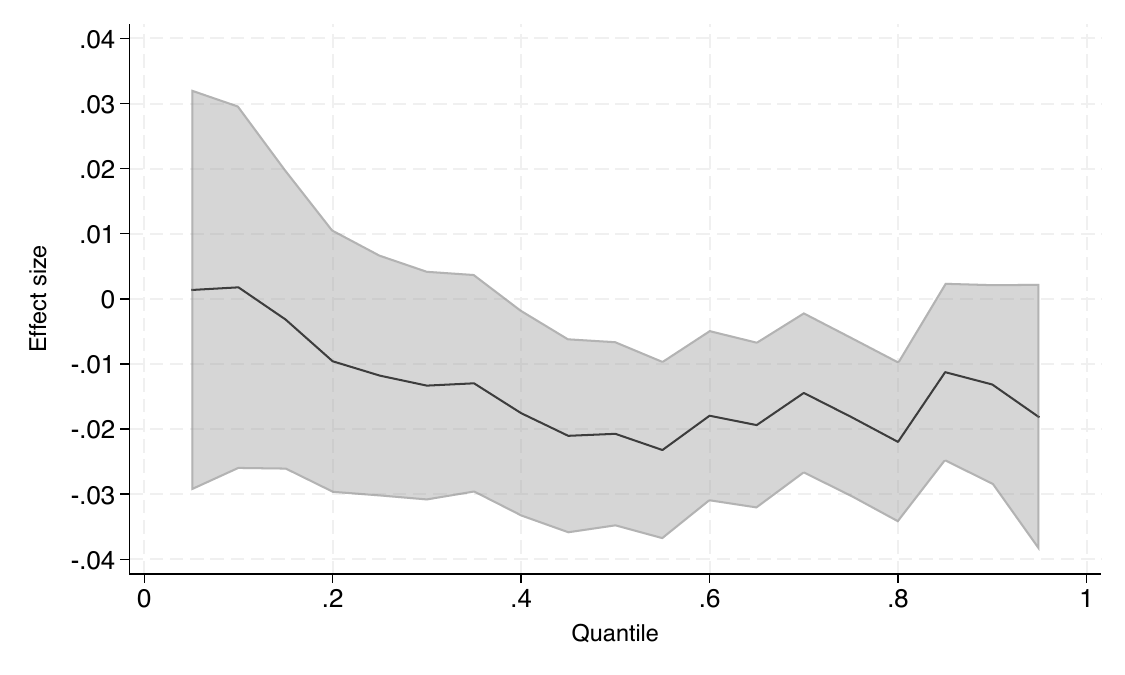}        
		\end{subfigure}
		\hfill		
		
		\begin{subfigure}{0.85\textwidth}
			\caption{Air Pollution Measured in $PM_{10}$}
			\includegraphics[width=\linewidth]{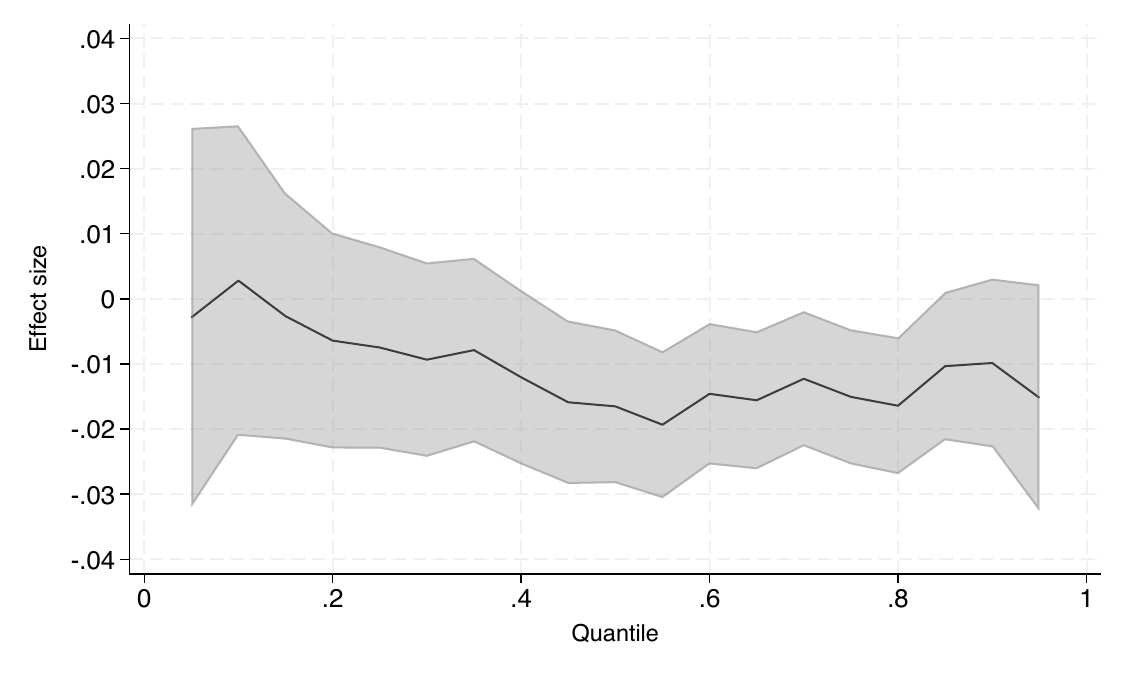}        
		\end{subfigure}
		
	\begin{minipage}{12cm}
		\vspace{0.2cm}
		\footnotesize{
			\textit{Notes}: The figures show the unconditional quantile treatment effects for every fifth percentile of the outcome distribution. In the top figure (a), air pollution is measured in $PM_{2.5}$, while in the bottom figure (b), it is measured in $PM_{10}$. The solid lines represent the quantile treatment effects and the gray areas around them are the 95\% confidence intervals. Both regressions include school and time fixed effects, weather controls (temperature, pressure, humidity), nonlinear weather controls (squared weather controls and interaction terms between those weather controls), and school-related control variables. The number of observations is 4,650.}
	\end{minipage}
		
\end{center}
\end{figure}

My results add to the above discussion by providing evidence from a country that has not yet been studied. This additional evidence may help to establish whether there is a potential explanation for the different gradient of air pollution effects found across studies. This explanation might relate to average air pollution values, the age of the individuals under study, or the country-specific educational setting. My results are presented in Figure \ref{quantile_graph} and Table \ref{quantile_table}. The graph shows the unconditional quantile treatment effects for every fifth percentile of the outcome distribution \citep{borgen2021new}. The table provides a more detailed analysis by identifying ten consecutive quantiles. Both analyses suggest that the higher the exam score, the greater the effects of air pollution. There are three potential and mutually nonexclusive explanations for this phenomenon.

The first explanation relates to the existence of ceiling and floor effects. These effects appear if subjects of the intervention are clustered at or near the upper (ceiling effect) or the lower (floor effect) limit of the dependent variable, such that the estimation of the treatment effects above or below certain values is impossible. If a high share of observations in the data exhibit the ceiling (floor) effects, researchers cannot observe the improving (deteriorating) effects of the treatment \citep{austin2002comparison}.
In this study, floor effects may appear. Each subject-specific exam consists of open-ended questions, but also single-choice tasks in which a student can simply be lucky and guess a correct answer.\footnote{In the exam in the national language (Polish) in 2022, 30\% of points were given for single-choice questions. The corresponding share for the math exam was 60\% and the modern foreign language (English) exam around 62\%.} This means that even the most unprepared pupils might score some points purely by chance and the lower bound of the actual exam scores might be way above zero.

The second explanation relates to the neuropsychological reactions of the human body. High-ability students may be more likely to perform at the upper edge of their ability, using their full brain capacity, which in turn does not leave any room to offset harmful air pollution effects \citep{zivin2020unintended}. In addition, it has been shown that under conditions of higher working memory load, individuals are more susceptible to distracting factors \citep{de2001role}. This may imply that performing more demanding tasks is more difficult in less hospitable conditions, for example with higher air pollution. It seems reasonable to assume that the difference in performance at the higher end of the exam score distribution results from how well the students solve these more difficult tasks, rather than how well they tackle the easy and medium-difficulty ones. The effects of changes in air pollution may then be limited to how these high cognitive load tasks are solved and therefore have a stronger effect on the final exam score among better-performing pupils \citep{krebs2024air}. Lastly, the work of \cite{kunn2019impact} shows that the negative impacts of air pollution in high-stakes situations increase when individuals operate under time pressure. Better-performing individuals may experience the pressure of time more strongly, since their aim is to correctly answer all the exam questions. Poor performers, on the other hand, may only aim to answer easier questions correctly, which leaves them with more time to rethink their answers. 

The third explanation, which is especially plausible in my setting where information on air quality is salient, relates to air pollution awareness. Thanks to the educational component of the ESA program, higher-ability pupils may be more conscious of the serious (physical and intellectual) consequences that air pollution has for humans. In addition, this increased awareness may also be reinforced be environment-related conversations which are more likely to be held in highly educated households. The combination of knowledge on the harmful effects of air pollution and information about the current air quality might result in elevated levels of stress and worry that external conditions out of the individual's control may influence the probability of a successful exam result.

To lend credibility to the above explanations, I first need to reject the hypothesis that better schools are simply located in areas with higher air pollution. Figure \ref{polution_vs_quality}	indicates that air pollution does not follow an increasing or decreasing trend across the whole spectrum of the schools' exam score distribution. Schools with the highest air pollution level are in the middle of the distribution. At the same time, schools with both very good and very bad exam results have the best air quality. This evidence suggests that the reason schools with better performers are more strongly affected by air pollution is not related to the quality of the air in these educational establishments.

Lastly, taking into account all evidence collected in recent research, I conclude that neither age nor general level of air pollution can explain a differential gradient of air pollution effects across different studies. Therefore, the factors that could potentially account for lower-performing individuals in some countries and better-performing ones in others being more strongly affected by bad air quality are most likely related to the country-specific educational setting or air pollution awareness combined with access to information.


\section{Benchmarking of Results}
\label{Benchmarking}

My results suggest that a 1 unit (1 $\mu g/m^{3}$) increase in the concentration of $PM_{2.5}$ leads to a 0.014 of a standard deviation decrease in the exam score. This translates into a 0.079 of a standard deviation decrease in the exam result for each standard deviation increase in $PM_{2.5}$. This effect is slightly greater than estimated by \cite{andersen2024air} and \cite{ebenstein2016long}. \cite{ebenstein2016long} find that a one standard deviation increase in $PM_{2.5}$ leads to a 0.038 of a standard deviation decrease in exam results, while \cite{andersen2024air} assess this effect to be 0.013--0.024 of a standard deviation, depending on the subject (see Table \ref{literature} for details).

Regarding the effects of $PM_{10}$, my findings imply that a 1 unit (1 $\mu g/m^{3}$) increase in $PM_{10}$ reduces the exam score by 0.011 of a standard deviation. This effect can also be interpreted as a decrease in the exam score by 0.07 of a standard deviation for each standard deviation increase in $PM_{10}$. The magnitude of this effect is sightly larger than the one obtained by \cite{carneiro2021effects}. In their paper, one standard deviation increase in $PM_{10}$ leads to a 0.05 of a standard deviation decrease in the exam score (see Table \ref{literature} for details).

To further assess the plausibility of my findings, a comparison of the effects based on $PM_{2.5}$ and $PM_{10}$ seems warranted. The more harmful effects of $PM_{2.5}$ compared to $PM_{10}$ found in my study are not surprising. The epidemiological research suggests that $PM_{2.5}$ is more dangerous for human health than $PM_{10}$, since these particles are very small (just 4\% of the diameter of a human hair) and can easily penetrate the body \citep{hei2018state}. Due to their extremely small size, they can move through nose and olfactory bulb directly into the brain. They can also cause indirect damage by triggering an inflammatory response \citep{bedi2021particle, underwood2017polluted}.

In the next step, I compare the effects of ambient air pollution found in my study to the effects of indoor air pollution (IAP) estimated in the literature. 
The concentration of indoor air pollution is predicted to be two to five times larger than outdoor air pollution, which leads to the conclusion that the two types of air pollution can be equally harmful \citep{vvallace1987total, world2010guidelines}. For example, in the area of London, \cite{metcalfe2025making} find that average daily concentration of $PM_{2.5}$ indoor is similar to the concentration measured outdoors; however, during typical occupancy hours (16.00--23.00), when individuals are most likely at home, this difference becomes much larger.\footnote{The daily average indoor air pollution concentration is 10.8 $\mu g/m^{3}$ and the outdoor concentration is 10.5 $\mu g/m^{3}$. During occupancy hours, the indoor pollution is 14.6 $\mu g/m^{3}$, while outdoor air pollution is 12.2 $\mu g/m^{3}$.} The main reason for this are household activities such as cooking, heating, smoking, or burning candles and incense. Since some of these actions are less likely to take place in schools, it is impossible to predict the indoor air pollution Polish pupils are exposed to during their exams. In addition, \cite{metcalfe2025making} show that although indoor and outdoor air pollution are correlated, the outdoor measures do not explain the indoor ones very well. Nevertheless, I can still compare my estimated effects of outdoor air pollution with the results of the paper by \cite{roth2016contemporaneous}, who directly assesses the effects of indoor air pollution. In the study, one standard deviation increase in $PM_{10}$ results in a 0.072 of a standard deviation decrease in the exam score. This finding is almost identical to my estimates of a 0.070 of a standard deviation decrease.

Overall, the negative effects of air pollution on cognitive performance in high-stakes settings estimated in my paper constitute the upper bound found in the literature to date, albeit with values still falling in the plausible range. There are at least two potential explanations for these disparities.

First, the difference in air pollution effect sizes between my paper and other studies may be attributed to the age of the affected individuals. While primary school leaving exams in Poland are taken by 14--15-year-olds, all other studies focus on upper secondary school exit exams or university entrance tests taken by 18--19-year-olds. At the same time, it has been established that children breathe through their mouths more often than adults \citep{bateson2008children}, have less efficient nasal filtering system \citep{bennett2008nasal}, and have a still underdeveloped immune system \citep{bearer1995children, schwartz2004air}. For all these reasons they can be more prone to the harmful effects of air pollution \citep{aithal2023air, bateson2008children}. Further research could validate whether these biological constraints also prevail in teenage years, and whether they might partly explain the difference in the effect sizes.

The alternative explanation for the slightly larger magnitude of air pollution effects on cognition found in my study compared to the existing literature relates to the precision of the air pollution measurements. All studies conducted to date use air pollution readings from devices located at least a few kilometers away from the potentially affected individuals. My paper is the first to use \textit{locatized} air pollution measurements collected directly at the location where pupils sit their exams. In light of this, I therefore assume that my air pollution measure is less likely to suffer from the measurement error problem and therefore provides more accurate estimates of the actual effect. The plausibility of this explanation is investigated in the following section.

\section{Spatial and Temporal Sensitivity of Air Pollution Effects}
\label{Reconciliation}

\subsection{Location of Air Pollution Measurement Sources}
\label{In_space}

The effects of air pollution on cognition in my study are estimated based on air pollution values recorded by the ESA devices installed in the schoolyard. I refer to this as the \textit{localized} air pollution measure, as both the devices and the observed individuals are at the same location. Since I also have data on air pollution values recorded by public air pollution measuring stations, I use this as an alternative measure of air quality. I refer to this as the \textit{distant} air pollution measure, as it is obtained at least a few kilometers away from the schools and the affected individuals. Due to a lack of other sources of measurement, most of the previous studies in the related literature used these public measuring stations as the primary source of air pollution information. The use of both measures and a comparison of the results obtained can be helpful in understanding whether and to what degree the remoteness of air pollution measurement matters.

The location of public air pollution measuring stations in Poland is presented
in Figure \ref{location_stations} and their characteristics are summarized in Table \ref{stations_charact}. There are 86 stations measuring the concentration of $PM_{2.5}$ and 168 stations measuring the concentration of $PM_{10}$. As these 86 $PM_{2.5}$ measuring stations are quite sparsely distributed across the country, in my comparison I focus only on $PM_{10}$ measurements, as well as AQI and a high pollution dummy variable calculated on the basis of $PM_{10}$.

The comparison of the effects of air pollution, depending on the location of air pollution measurement source, is presented in Table \ref{main_space}. In column 1, the effects from the main specification with the localized air pollution measure are reported.\footnote{These results are the same as those presented in column 4 of Table \ref{main_results}.} In columns 2--4, distant air pollution measures are used. In column 2, the air pollution measures come from the closest measuring station to the given school. Columns 3 and 4 present the readings from the two and three closest measuring stations, respectively. In these two specifications, the average air pollution values are calculated using inverse distance weighting (IDW) \citep{bondy2020crime, currie2005air, sager2019estimating}. The weighting procedure is also conducted for the weather controls which include temperature, precipitation, and cloud cover. Further, all specifications include the same set of school time-varying characteristics. The respective air pollution measurements -- $PM_{10}$, AQI, and the high pollution dummy variable based on $PM_{10}$ -- are presented in consecutive rows of the table.

The effects estimated using the \textit{distant} air pollution measures, presented in columns 2--4, are insignificant and their magnitude is close to zero. The discrepancy between these effects and the significant, negative impacts estimated using the \textit{localized} air pollution measures (column 1) can be attributed to at least three factors.

The first source of discrepancy between the results obtained using \textit{localized} and \textit{distant} air pollution measures is the distance to the potentially affected individuals. While ESA devices are located directly in the schoolyard, the mean distance from the schools to the closest air pollution measuring station is around 16 km. When two and three measuring stations are considered, this distance increases to around 21 and 26 km, respectively. This remotedness of the \textit{distant} air pollution measure is a potential reason for the mismeasurement of air pollution exam takers are exposed to and thus leads to the misestimation of the effects.

This explanation complies with the theoretical predictions under the classical errors-in-variables (CEV) assumption. According to this assumption, the measurement error is unrelated to the true unobserved explanatory variable, while at the same time correlating with the observed erroneous measure. This leads to a biased and inconsistent estimator, which is always closer to zero than the true population parameter. This bias is referred to as the attenuation bias. In practice, it means that the effects obtained with the mismeasured variable are more likely to be close to zero in magnitude. In my analysis, this is precisely the consequence of using the \textit{distant} air pollution measure -- effects estimated based on this metric lean toward zero, compared to the negative effects obtained with the presumably more accurate \textit{localized} metric.

An alternative hypothesis for explaining the difference in the effects between the \textit{localized} and \textit{distant} measures of air pollution relates to the types of measuring devices. As noted earlier, for the ESA program, low-cost but high-quality measuring devices are installed. At the same time, public air pollution measuring stations mostly use the gravimetric (reference) method or a similarly credible automatic method, as recommended by relevant EU directives.\footnote{The current literature seems to use a variety of different measurement methods but the lack of in-depth description of these methods prevents adequate assessment of their credibility. For example, \cite{roth2016contemporaneous}  uses low-cost yet advanced portable air pollution measurement devices, while \cite{andersen2024air} uses a high-resolution modeling system to obtain air pollution values on a 1 km x 1 km grid.} 

Nevertheless, I believe that in my study the differences in the measuring technique are unlikely to have contributed to the distinct estimation results. As Figure \ref{int_vs_ext} shows, both air pollution measures are highly correlated with the pairwise correlation coefficient of 0.849.\footnote{The air pollution values measured by public measuring stations are, on average, slightly lower in the fall/winter season and slightly higher in the spring/summer season compared to air pollution values recorded by the ESA devices.}
Moreover, since the fixed effects model uses the within variation, the absolute values of air pollution are not as important as the differences between them within a unit of observation.\footnote{As shown in column 1 of Table \ref{main_space}, the ESA devices record mean air pollution measured in AQI of around nine units, while public air pollution measuring stations record a value of around 20 units (columns 2--4).} As the within-unit deviations from the mean exhibit similar values irrespective of the air pollution metrics, I claim that both \textit{localized} and \textit{distant} air pollution measures should capture enough true within-unit variation to make uncovering the effects of air pollution on cognition possible.

The third potential reason explaining the difference in the results between the \textit{localized} and \textit{distant} air pollution measurement is a difference in measurement objectives. While the ESA devices aim to measure air pollution individuals are exposed to during their time at school, the goal of public measuring stations is to capture air pollution in the most representative manner. According to the Regulation of the Minister of Climate and Environment of December 11, 2020 on Assessing the Levels of Substances in the Air, public air pollution measuring stations should be located such that they provide data from areas with the highest levels of substances in the air and are representative for an area of several square kilometers in the case of urban background measurements.\footnote{Background stations in urban areas are the most popular type of air pollution measuring station in Poland -- see Table {\ref{stations_charact}.}} Further, the air flow around the intake should not be restricted by any obstacles and the intake inlet should be located a few meters away from buildings, balconies, trees, and other obstacles. These specifications ensure that measurements from public air pollution measuring stations present a representative picture of air pollution in Poland. However, such a nationwide representative picture may not be the most suitable measure of air pollution in particular locations such as the schoolyard.

All in all, the lack of significant effects of air pollution on cognition when using the \textit{distant} air pollution metric indicates the importance of precise air quality measurement. In my study, only \textit{localized} air pollution ensures credible estimates unlikely to be contaminated by the attenuation bias. \textit{Distant} air pollution measures, on the other hand, induce bias and prevent a rigorous estimation of the effects. This conclusion does not undermine the findings of other papers, which used \textit{distant} air pollution measures and found significant, negative impacts. Counterwise, my study provides the first evidence that the effects reported in current studies might be slightly underestimated. The reason for this is that the average distance between exam locations and public air pollution measuring stations in my setting is much larger than those utilized in the literature.\footnote{For example, \cite{carneiro2021effects} use the readings from the measuring stations located up to 10 km from the municipality in which the exams are conducted. \cite{ebenstein2016long}, on the other hand, consider the readings from the measuring stations within a maximum of 2.5 km from the exam location.} The smaller difference in distance between potentially affected individuals and public air pollution measuring stations in other papers might be why they still estimated significant and negative effects.

Lastly, to further unpack the above conclusion, I investigate whether considering only those schools located up to a given distance from the closest air pollution measuring station changes the estimates obtained. Figure \ref{effect_by_disatnce}	shows the results of this exercise. First, I restrict the sample of schools to those located up to 15 km from the closest air pollution measuring station. Further, the subsample of schools is reduced to those located up to 10 km, 5 km, 3 km, 2 km, and 1 km away. The results indicate that the effects of air pollution are still insignificant and close to zero, irrespective of the distance.\footnote{When air pollution is measured in $PM_{10}$ and as AQI, and if the sample is limited to schools located within 5 km of the closest public air pollution measuring station, the estimated coefficients are positive and significant at the 5\% level. However, as this effect is not confirmed when using a binary measure of air pollution or with different distance values, I prefer not to overinterpret this result.} One explanation for this finding is that by limiting the maximum distance, I also reduce the sample size and its representativeness. For schools located up to 15 km from the closest air pollution measuring station, the sample size is reduced by almost 50\% (N = 2,542), while in case of a 10 km threshold, the sample is almost 70\% (N = 1,748) smaller.\footnote{The sample size for the schools located up to 5 km away is 1,102, for those up to 3 km away it is 707, up to 2 km it is 436, and up to 1 km 117 schools.} This smaller sample size leads to the increase in the confidence intervals and reduced precision. 

\subsection{Time Window of Air Pollution Measurement}
\label{In_time}

The timing of the air pollution measurement may also affect the magnitude of the
estimated effects of air quality on cognitive performance. Figure \ref{over_day_time} supports this statement by showing the evolution of air pollution over the course of the days on which the exams took place. Although there are significant differences in levels between examination dates, there is a common hourly pattern. Air pollution starts to increase in the early evening and remains at its highest level throughout the night. In the early morning, air pollution decreases and the lowest levels are recorded around late afternoon.

To estimate the effects of air quality on cognition, most previous studies use daily average air pollution values \citep{bondy2020crime, carneiro2021effects, ebenstein2016long, sager2019estimating}. One reason they follow this strategy is to avoid measurement error. The greater the variability in air pollution over the day, the larger the measurement error if only values from a restricted time window are considered. The second reason is that besides the contemporaneous effects during the examination, air pollution children are exposed to some time before the exam may induce lingering effects.

I follow the current literature and use daily average air pollution values in my main specification. Given that the primary school leaving exams take place in the morning (9:00--11:00 a.m.), hourly air pollution from the last 24 hours is used to calculate the daily average.\footnote{This means that the hourly air pollution 22 hours before the examination (11:00--09:00 a.m.) and for the 2 hours of the examination (9:00--11:00 a.m.) is used to calculate the daily average air pollution values.} The effects estimated using this measure are reported in the main Table  \ref{main_results}, and again in column 1 of Table \ref{main_time}. In other columns in this table, the time window used to calculate the daily average air pollution is reduced. In column 2, the last 12 hours (10 hours before examination and the 2-hour exam time window) is considered. In column 3, the last 4 hours (2 hours before the exam and the 2 hours of the examination) are used. Lastly, in column 4, only the air pollution measured during the examination (2-hour time window) is used.

The main conclusion from the comparison of the effects in Table \ref{main_time} is that the effects of air pollution measured shortly before the examination time are smaller and less significantly estimated than the effects from a longer time period before the exam. There may be two reasons for this result. First, using fewer air pollution values from a more limited time window may result in larger measurement error and consequently greater attenuation bias. That is, the daily average calculated based on a 2-hour time window is more prone to outlying observations than the daily average calculated based on 4-, 12-, and 24-hour time windows. Second, smaller effects estimated in a shorter time window may be explained by the existence of lingering effects of air pollution. All the exams take place in the morning (9:00--11:00 a.m.) when air pollution is the lowest. Nevertheless, primary school children usually live very close to their schools (85\% up to 3 km away) and are therefore affected by similar air quality during exam preparation or while resting at night.\footnote{In urban areas, the average distance from place of residence to the closest primary school is 0.8 km, while in rural areas, this distance is 2.1 km. Almost 100\% of children in urban areas live less than 3 km away from the closest school, with 77.5\% living less than 1 km away. In rural areas, these percentages are 75.0\% and 31.3\%, respectively (Statistics Poland, 2016).} The lingering effects from some time before the exams may matter more than the air pollution they are exposed to at the exact time of the examination.

\begin{figure}
	\begin{center}
		
		\caption{Effects of Air Pollution \\ One Day Before and One Day After the Exams}
		\vspace{0.1cm} 
		\label{lead_lag}
		
		\begin{subfigure}{0.85\textwidth}
			\caption{Air Pollution Measured in $PM_{2.5}$}
			\includegraphics[width=\linewidth]{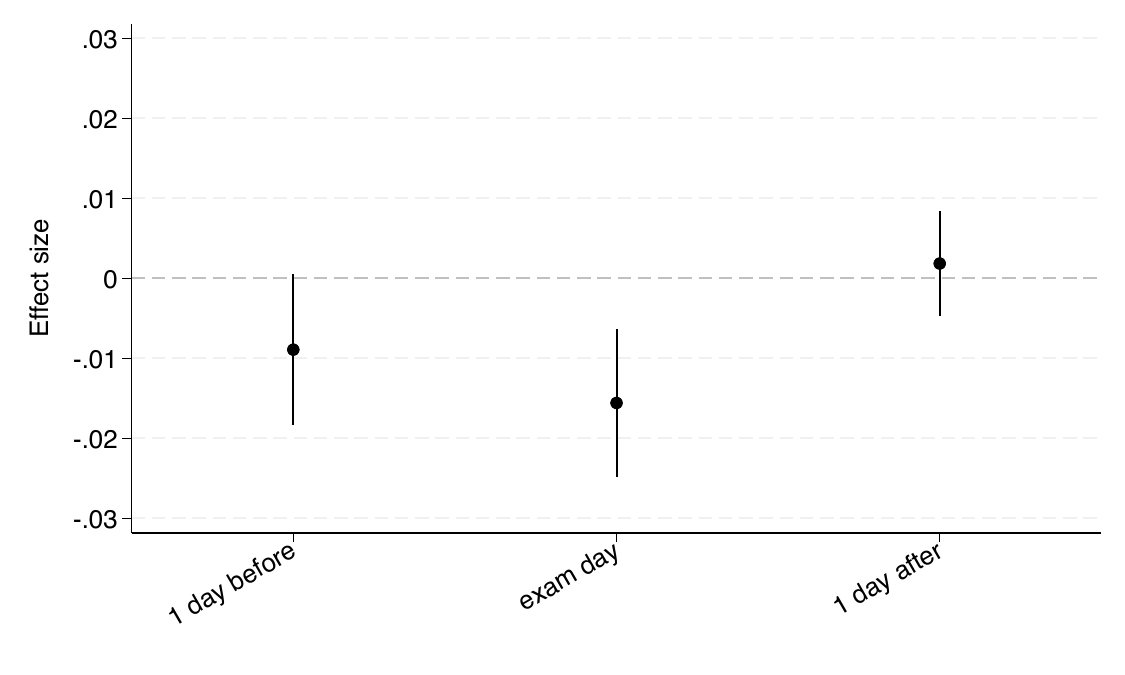}        
		\end{subfigure}
		\hfill		
		
		\begin{subfigure}{0.85\textwidth}
			\caption{Air Pollution Measured in $PM_{10}$}
			\includegraphics[width=\linewidth]{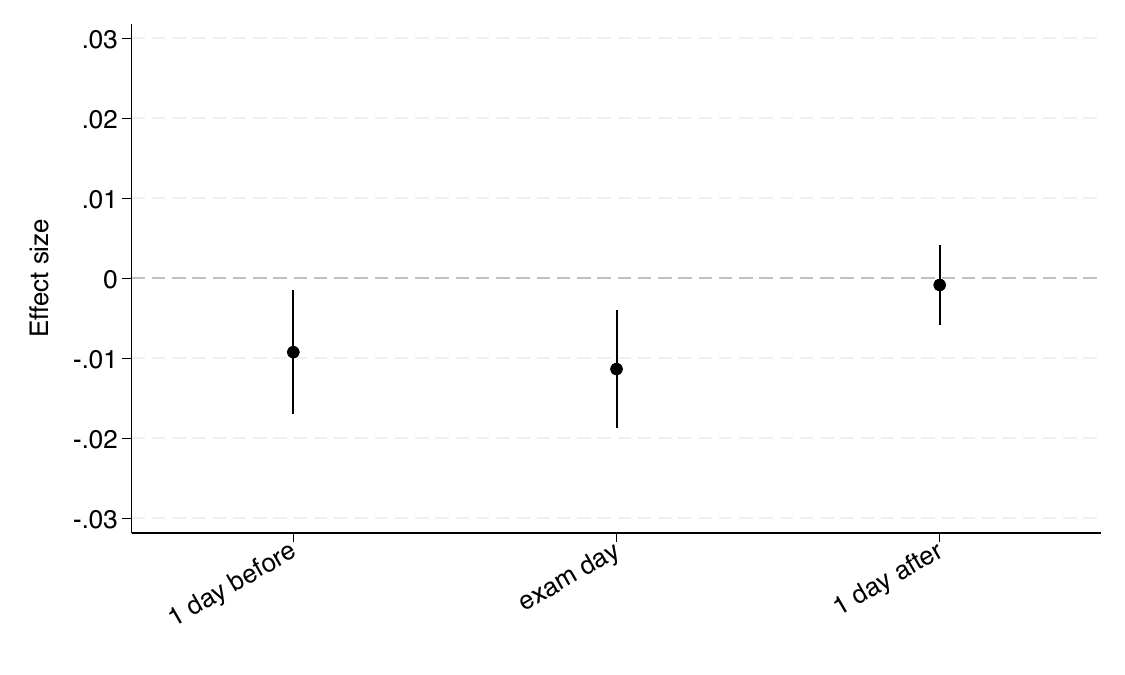}        
		\end{subfigure}
		
	\begin{minipage}{17cm}
		\vspace{0.09cm}
		\footnotesize{
			\textit{Notes}: The figures show the effects of lagged, contemporaneous, and lead air pollution on the standardized exam results together with 95\% confidence intervals. There are three coefficients on each graph from the same regression. The first coefficient represents the effects of lagged air pollution measured one day before the exams. The second coefficient represents the contemporaneous effects of air pollution measured on the day of the exams. The third coefficient represents the lead air pollution measured one day after the exams. In the top figure (a), air pollution is measured in $PM_{2.5}$, while in the bottom figure (b), it is measured in $PM_{10}$. All regressions include school and time fixed effects, weather controls (temperature, pressure, humidity), nonlinear weather controls (squared weather controls and interaction terms between these weather controls), and school-related control variables. The number of observations is 4,650.}
	\end{minipage}
		
\end{center}
\end{figure}

To further investigate the lingering effects hypothesis, I examine whether the daily average air pollution from one day before the exam also affects the exam score. At the same time, I test the placebo effects by including the daily average air pollution from one day after the exam.\footnote{I consider air pollution values measured just one day before the exam, as this is when the chances of the observed pupils being in school or at home nearby are the highest. Two (three) days before the examination would include Saturday (and Sunday), when students may be residing further away from school. Indeed, activities such as traveling are even recommended by school psychologists to reduce the stress students experience before their first important examination.} The expectation is that air quality one day before the exam may have significant and negative effects on the exam results but these effects are expected to be smaller than those measured on the day of the exam. Further, the effects of air pollution from one day after the exam should not induce significant impact. These predictions are confirmed by Figure \ref{lead_lag}. Air pollution on the day of the exam has the strongest impact on the exam results. Air quality one day before the exam also induces negative effects, although these effects are smaller and estimated to be less significant. Lastly, the effects of air pollution one day after the exam are estimated to be close to zero.

Taken together, the above findings provide additional support for the importance of a meticulous air pollution measure. Further, not only the air quality pupils are exposed to during the examination, but also that experienced a few hours before the test has a significant impact on the cognitive performance of the exam takers. This shows that besides the contemporaneous effects, the lingering effects of air pollution should be taken into account when assessing the overall impact of air quality on human cognition.


\section{Discussion}
\label{discussion}

This section puts the estimated effects into a broader perspective by assessing their economic significance for individuals and for society as a whole. 
As explained in Section \ref{Exams}, primary school leaving exam results influence the probability of admission to different secondary schools. A higher score increases the likelihood of being placed in more prestigious, better-equipped and academically selective institutions. A decline in exam score caused by exposure to air pollution may therefore see students redirected toward lower-quality alternatives. Such a change in school placement can result in a change in the individual's human capital development trajectory and induce long-run consequences for educational attainment and labor market outcomes such as earnings \citep{chetty_measuring_2014, jackson}.

Because the primary school leaving exams are graded in integers, air pollution would need to alter the exam score by at least one full point to potentially affect admission probability. According to my estimates, a one-point change in exam score would require a change in air pollution of 72 units of $PM_{2.5}$ or 91 units of $PM_{10}$. By way of comparison, the variation in air pollution between schools on exam days amounts to around 4.3 units of $PM_{2.5}$ and 4.9 units of $PM_{10}$. This means that the exogenous changes in air quality on exam days are unlikely to influence the secondary school admission outcome in my context.

The above conclusion is not entirely in line with \cite{brehm2022low}, who find
that a reduction in air pollution can improve transitions within the education system. However, the lack of effects in my setting may stem from a specific institutional feature -- the integer-based grading scheme. This characteristic creates a minimum threshold below which pollution-induced changes in performance have no impact on admission outcomes. Due to this specific aspect, my conclusion may not be externally valid. The grading scheme may alter whether or not air pollution impacts admission probability and consequently further educational outcomes.
 
The effects of air pollution on students' cognitive performance in high-stakes educational situations may also extend to other cognitively demanding environments, such as the workplace. My findings suggest that environmental shocks can significantly impair an individual's ability to perform complex mental tasks, even with strong incentives and optimal motivation. Performance in high-stakes exams can be seen as analogous to meeting deadlines, multitasking, problem-solving, strategic thinking, or accurate and fast decision-making. Achievements in all these domains, especially in white-collar jobs, might affect worker's efficiency and productivity. This conclusion is supported by recent empirical research showing that air pollution can indeed adversely affect the productivity of high-skill workers \citep{archsmith2018air, chang2019effect}. At the firm level, these individual impairments may aggregate into reduced output, lower profitability, and weaker long-term growth potential.

The short-term effects of air pollution on students' cognitive performance in high-stakes educational settings also raise broader concerns about the implications for human capital formation. Cognitive skills developed during childhood and adolescence are critical in the lifelong process of human capital accumulation. They are also known to predict educational achievement, labor market success, and earnings \citep{cunha2010estimating, hanushek2015returns}. If environmental shocks such as air pollution impair performance at critical educational junctures, they may also reduce the overall quality of the acquired skills. Repeated exposure to such shocks increases the risk of human capital deficits. Recent empirical studies have established causal links between environmental conditions and human capital formation \citep{graff2013environment, park2017temperature}. My findings add to this growing literature by providing evidence that short-term exposure to pollution might disrupt both the development and deployment of cognitive skills -- both of which are important foundations of human capital.

Lastly, the implications of impaired cognitive performance due to air pollution may extend beyond the individual or firm level, potentially affecting aggregate economic outcomes such as long-run GDP growth. Recent growth models foreground human capital suggesting that countries with a higher aggregate level of skills innovate at a faster rate \citep{hanushek2010high}. At the same time, these models emphasize that the quality of the acquired skills and knowledge -- rather than the quantity of schooling -- should be taken as a key measure of a country's human capital stock \citep{hanushek2012schooling, hanushek2015returns}. Empirical evidence demonstrates that international differences in student achievement, measured by standardized assessments such as PISA or TIMSS, explain a substantial share of the variation in long-run GDP growth rates across countries \citep{hanushek2011economics, hanushek2012better}.\footnote{PISA (Programe for International Student Assessment) is an international study aimed at obtaining comparable data on the skills of students aged 15 and over. TIMSS (Trends in International Mathematics and Science Study) is a worldwide series of international assessments of the mathematics and science knowledge of students.} Therefore, if environmental factors like air pollution reduce students' performance in high-stakes exams, thereby lowering the stock of human capital, then persistent or widespread exposure to air pollution may also have long-term macroeconomic consequences reflected in slower economic growth.

The above findings have several important policy implications. At the individual level, reducing exposure to air pollution might be a cost-effective strategy for enhancing productivity and supporting human capital development. At the societal level, improving air quality could lead to increases in national productivity and contribute to sustainable economic growth. Targeted environmental interventions -- such as installing air filters or improving ventilation systems -- therefore offer benefits that extend beyond better health outcomes. These results underscore the need to integrate environmental policy into the educational and labor market policy framework.


\section{Conclusion}

This paper analyzes the short-term effects of air pollution on the cognitive performance of students in a high-stakes setting. Cognition is measured based on the results of primary school leaving exams taken by 14--15-year-old pupils in Poland. I use unique data containing information on the concentration of particulate matter continuously measured directly in the schoolyard for three consecutive years (2022--2024). The school-level panel data make it possible to control for school and time fixed effects, as well as weather- and time-varying school-related controls. The results obtained are benchmarked against the findings in the current literature. Lastly, I compare my estimates to those obtained using a \textit{distant} air pollution measurement, drawn from public air pollution measuring stations, and to the results obtained using a modified time window in which air pollution is measured. 

The main results indicate that a 10 unit ($\mu g/m^{3}$) increase in $PM_{2.5}$ and $PM_{10}$ leads to a decrease in exam results of around 0.14 and 0.11 of a standard deviation, respectively. Schools with poor air quality, defined by the WHO as the concentration of $PM_{2.5}$ above 15 $\mu g/m^{3}$, experience around a 0.20 standard deviation decrease in their exam results compared to schools without such conditions. These results constitute the upper bound found in the literature to date. My study also provides evidence that the harmful effects of air pollution are larger at the higher end of the exam score distribution, thus affecting better-performing students more than lower-performing ones. Further, the results obtained using \textit{distant} air pollution measures drawn from public air pollution measuring stations are smaller in magnitude and not statistically significant. This points to the importance of precise air pollution measures for the estimation of causal effects. Lastly, modifying the time window in which air pollution is measured indicates that the effects of air pollution are not only contemporaneous, but also linger from several hours before the examination.

Although my results are robust to a wide range of robustness exercises and placebo tests, my analysis has several important limitations. First, despite the inclusion of a rich set of control variables, there may still be some time-varying and unobserved factors affecting both the level of air pollution and exam results. Second, the aggregation of data at the school level does not allow me to control for students' characteristics and therefore to investigate heterogeneity at the individual level. Third, I lack data-driven explanations for the stronger effects of air pollution at the higher end of the exam score distribution. These aspects could be the focus of future research within the related literature. 

Taken together, the findings of my study underscore the substantial impacts that environmental conditions -- especially air quality -- can have on cognitive performance in high-stakes settings. 
By leveraging highly \textit{localized} air pollution data, I demonstrate that the effects of air pollution may be larger than previously estimated using \textit{distant} air pollution sources. This suggests that the cognitive costs of poor air quality may have been systematically underestimated in earlier research. 
My results also reveal that the influence of air pollution is not limited to the duration of the cognitive effort itself, but extends to the period preceding the mentally demanding activity. This challenges the notion that short-lived interventions alone are sufficient to mitigate the detrimental effects of air pollution. 
More broadly, my study contributes to a growing literature showing that the consequences of air pollution extend beyond physical health, also affecting non-health-related outcomes such as education attainment, productivity, and human capital formation.

	
	\newpage
	\bibliography{biblio.bib}
	\newpage
	
	
	 \newpage
	 \section*{Figures and Tables}
	 \label{Figures_and_Tables}
	 

\clearpage

\clearpage

\clearpage

\clearpage

\clearpage

\clearpage

\begin{table}
	\begin{center}	
		\caption{Summary Statistics}
			\label{sum_stats}
			\begin{tabular}{lcccccc}
					\toprule \toprule
	&  \multirow{2}{*}{Mean} & Standard & Number of \\ 
	&   & deviation & observations \\ 
	\cline{2-4}
	&  (1)       & (2)     & (3) \\

	\midrule
	\vspace{0.1cm}
			
	\textit{Panel A: Outcomes} \\	
	\vspace{0.15cm}
	\hspace{2mm} Pooled standardized exam result&       0.00&        1.00&        4,650\\
	\vspace{0.05cm}
	\hspace{2mm} Polish exam result&       61.60&        9.15&        1,567\\
	\vspace{0.05cm}
	\hspace{2mm} Math exam result&       50.45&       12.94&        1,568\\
	\vspace{0.2cm}
	\hspace{2mm} English exam result &       63.24&       12.32&        1,515\\
							
	\textit{Panel B: Air pollution} \\	
	\vspace{0.05cm}
	\hspace{2mm} $PM_{2.5}$ ($\mu g/m^{3}$)&        7.74&        5.62&        4,650\\
	\vspace{0.05cm}
	\hspace{2mm} $PM_{10}$  ($\mu g/m^{3}$)&        9.33&        6.39&        4,650\\
	\vspace{0.05cm}
	\hspace{2mm} AQI&        8.70&        5.87&        4,650\\
	\vspace{0.05cm}
	\hspace{2mm} High pollution ($PM_{2.5}$)&        0.13&        0.33&        4,650\\
	\vspace{0.2cm}
	\hspace{2mm} High pollution ($PM_{10}$)&        0.08&        0.27&        4,650\\
				
	\textit{Panel C: Weather-related control variables} \\	
	\vspace{0.05cm}
	\hspace{2mm} Temperature (\degree C) &       16.81&        3.54&        4,650\\
	\vspace{0.05cm}
	\hspace{2mm} Pressure (hPa) &      994.46&       15.69&        4,650\\
	\vspace{0.2cm}
	\hspace{2mm} Humidity (\%)     &       60.02&       18.49&        4,650\\
				
	\textit{Panel D: School-related control variables} \\	
	\vspace{0.05cm}
	\hspace{2mm} Number of exam takers &       34.80&       30.90&        4,650\\
	\vspace{0.05cm}
	\hspace{2mm} Number of 8th grade students &       38.32&       33.14&        4,650\\
	\vspace{0.05cm}
	\hspace{2mm} Number of pupils &      297.17&      201.87&        4,650\\
	\vspace{0.05cm} 
	\hspace{2mm} Share of female 8th grade students &       47.93&       12.19&        4,650\\
	\hspace{2mm} Share of female pupils &       48.30&        4.54&        4,650\\
	
	\bottomrule
		\end{tabular}
			\begin{minipage}{15cm}
				\vspace{0.2cm}
					\footnotesize{
					\textit{Notes}: The table presents the summary statistics for the analyzed sample. Panel A shows the average subject-specific exam results and the pooled standardized exam result. The subject- specific exam results are measured on a scale from 0 to 100 points. The pooled standardized exam result has a mean of zero and a standard deviation of one for the whole sample. Panel B shows the average air pollution values. First, air pollution is measured as the concentration of $PM_{2.5}$. Second, air pollution is measured as the concentration of $PM_{10}$. Third, Air Quality Index (AQI) is calculated based on the $PM_{10}$ values. Further, the high pollution dummy based on $PM_{2.5}$ values is defined as 1 if the daily air pollution concentration is equal to or greater than 15 $\mu g/m^{3}$, and to 0 otherwise. Lastly, the high pollution dummy based on $PM_{10}$ values is defined as 1 if the daily air pollution concentration is equal to or greater than 20 $\mu g/m^{3}$, and to 0 otherwise. Panel C shows weather-related control variables and panel D shows school-related control variables. The number of exam takers indicates the number of students taking the school leaving exams and may be lower than the number of 8th grade students for the reasons outlined in the Data and Empirical Strategy section. The number of pupils indicates the total number of students in a school. Observations are from the school years 2021/22--2023/24. The total number of observations is 4,650. The number of observed schools is 788.}
			\end{minipage}
		\end{center}
	\end{table}
	

\clearpage

\begin{table}
		\begin{center}
			\caption{Effects of Air Pollution on Exam Results}
			\label{main_results}
			\begin{tabular}{lccccccccccccc}
				\toprule \toprule
				& \multicolumn{4}{c}{Standardized Exam Result} \\
				\cline{2-5}
				&  (1) & (2) & (3) & (4) \\
				\midrule
				
	$PM_{2.5}$ ($\mu g/m^{3}$)&      -0.015$^{***}$&      -0.014$^{***}$&      -0.015$^{***}$&      -0.014$^{***}$\\
	\vspace{0.3cm}
	\hspace{3mm}&     (0.004)   &     (0.004)   &     (0.004)   &     (0.004)   \\
			
	$PM_{10}$  ($\mu g/m^{3}$)&      -0.011$^{***}$&      -0.011$^{***}$&      -0.012$^{***}$&      -0.011$^{***}$\\
	\vspace{0.3cm}
	\hspace{3mm}&     (0.003)   &     (0.003)   &     (0.003)   &     (0.003)   \\
				
	AQI&      -0.012$^{***}$&      -0.012$^{***}$&      -0.013$^{***}$&      -0.012$^{***}$\\
	\vspace{0.3cm} 
	\hspace{3mm}&     (0.003)   &     (0.004)   &     (0.004)   &     (0.004)   \\

	High pollution ($PM_{2.5}$) &      -0.202$^{***}$&      -0.206$^{***}$&      -0.213$^{***}$&      -0.206$^{***}$\\
	\vspace{0.3cm} 
	\hspace{3mm}  &     (0.036)   &     (0.037)   &     (0.037)   &     (0.037)   \\
				
	High pollution ($PM_{10}$)&      -0.188$^{***}$&      -0.186$^{***}$&      -0.186$^{***}$&      -0.181$^{***}$\\
	\vspace{0.3cm} 
	\hspace{3mm} &     (0.040)   &     (0.042)   &     (0.042)   &     (0.041)   \\

	\vspace{0.3cm} 
	Number of observations & 4,650 & 4,650 & 4,650 & 4,650 \\
				
	Weather controls & No & Yes & Yes & Yes \\
	Non-linear weather controls & No & No & Yes & Yes \\
	School controls & No & No & No & Yes \\
				
	\bottomrule
		\end{tabular}
			\begin{minipage}{14cm}
				\vspace{0.2cm}
					\footnotesize{
					\textit{Notes}: The table shows the estimated effects of air pollution on the standardized exam results. Each coefficient is from a separate regression. In the first and second rows, air pollution is measured as the concentration of $PM_{2.5}$ and $PM_{10}$, respectively. In the third row, the Air Quality Index (AQI) is calculated based on the $PM_{10}$ values. In the fourth row, the high pollution dummy is based on $PM_{2.5}$ values and defined as 1 if the daily air pollution concentration is equal to or greater than 15 $\mu g/m^{3}$, and to 0 otherwise. In the last row, the high pollution dummy is based on $PM_{10}$ values and defined as 1 if the daily air pollution concentration is equal to or greater than 20 $\mu g/m^{3}$, and to 0 otherwise. The specification in column 1 includes only school and time fixed effects. The specification in column 2 also includes weather controls (temperature, pressure, humidity). The specification in column 3 is enriched by the nonlinear weather controls (squared weather controls and interaction terms between these weather controls). The specification in column 4 is the preferred specification, which also includes school-related control variables. All estimations are based on 4,650 observations and the sample mean of the outcome variable is 0.000. Robust standard errors (presented in parentheses) allow for clustering at the school level. *** p$<$0.01, ** p$<$0.05, * p$<$0.1.}
			\end{minipage}
		\end{center}
	\end{table}


\clearpage

\begin{landscape}

\begin{table}
		\begin{center}
			\caption{Robustness Checks: \\ Effects of Air Pollution on Exam Results}
			\label{robustness}
			\begin{tabular}{lccccccccccccc}
				\toprule \toprule
				& & & & \multicolumn{2}{c}{Sample Selection} & & \multicolumn{2}{c}{COVID-19 Severity}\\
				\cline{5-6}
				\cline{8-9}
				& SE Clust. & Temp. & & Public & Weighted & & Number & Positive \\
				& at Reg. Level & in Bins & & Schools & Sample & & of Cases & Test Rate \\
				\cline{2-9}
				& (1) & (2) & & (3) & (4) & & (5) & (6) \\
				\midrule
				
	$PM_{2.5}$ ($\mu g/m^{3}$)&      -0.014$^{***}$&      -0.014$^{***}$&   &   -0.015$^{***}$  &      -0.013$^{*}$  &&      -0.014$^{***}$&      -0.015$^{***}$\\
	\vspace{0.3cm}
	\hspace{3mm} &     (0.004)   &     (0.004)   &  &   (0.004) &     (0.008)   &&     (0.004)   &     (0.004)   \\
			
	$PM_{10}$  ($\mu g/m^{3}$)&      -0.011$^{***}$&      -0.012$^{***}$&   &   -0.012$^{***}$ &      -0.012$^{*}$  &&      -0.011$^{***}$&      -0.011$^{***}$\\
	\vspace{0.3cm}
	\hspace{3mm}&     (0.003)   &     (0.003)   & &    (0.003)   &     (0.006)  & &     (0.003)   &     (0.003)   \\
				
	AQI &      -0.012$^{***}$&      -0.013$^{***}$&   &   -0.013$^{***}$&      -0.013$^{*}$ &&      -0.012$^{***}$&      -0.012$^{***}$\\
	\vspace{0.3cm} 
	\hspace{3mm} &     (0.004)   &     (0.004)   &    & (0.004)   &     (0.007)   &&     (0.004)   &     (0.003)   \\

	High pollution ($PM_{2.5}$) &      -0.206$^{***}$&      -0.212$^{***}$& &     -0.224$^{***}$ &      -0.133$^{**}$ &&      -0.207$^{***}$&      -0.213$^{***}$\\
	\vspace{0.3cm} 
	\hspace{3mm} &     (0.037)   &     (0.037)   & &    (0.037)   &     (0.064)   & &     (0.036)   &     (0.036)   \\
				
	High pollution ($PM_{10}$) &      -0.181$^{***}$&      -0.186$^{***}$&   &   -0.190$^{***}$  &      -0.127$^{*}$  &&      -0.175$^{***}$&      -0.179$^{***}$\\
	\vspace{0.3cm} 
	\hspace{3mm} &     (0.042)   &     (0.042)   &  &   (0.042)   &     (0.073)   &&     (0.041)   &     (0.042)   \\
	
	Number of observations  &        4,650   &        4,650   &   &     4,486 & 4,650   &&        4,650 &        4,650 \\
		
	\bottomrule
		\end{tabular}
			\begin{minipage}{19cm}
				\vspace{0.2cm}
					\footnotesize{
					\textit{Notes}: The table shows the robustness checks for the regression of standardized exam results on air pollution. In the first and second rows, air pollution is measured as the concentration of $PM_{2.5}$ and $PM_{10}$, respectively. In the third row, the Air Quality Index (AQI) is calculated based on the $PM_{10}$ values. In the fourth row, the high pollution dummy is based on $PM_{2.5}$ values and defined as 1 if the daily air pollution concentration is equal to or greater than 15 $\mu g/m^{3}$, and to 0 otherwise. In the last row, the high pollution dummy is based on $PM_{10}$ values and defined as 1 if the daily air pollution concentration is equal to or greater than 20 $\mu g/m^{3}$, and to 0 otherwise. In column 1, standard errors are clustered at the municipality level. In column 2, the temperature is measured in bins of the size 2\degree C. In columns 3--4, the sample selection issues are considered. In column 3, the sample of schools is limited to public schools only. In column 4, the weighting is incorporated to adjust the sample to the known population characteristics. In columns 5--6, the potentially confounding effects of the COVID-19 pandemic are considered. The specification in column 5 controls for the average number of COVID-19 cases [per 100,000 inhabitants] in each month of the final school year in the county where a given school is located. The specification in column 6 controls for the positive COVID-19 test rate in each month of the final school year in the county where a given school is located. Robust standard errors are presented in parentheses. *** p$<$0.01, ** p$<$0.05, * p$<$0.1.}
			\end{minipage}
		\end{center}
	\end{table}
	
\end{landscape}


\clearpage

\begin{table}
		\begin{center}
			\caption{Effects of Air Pollution on Exam Results}
			\vspace{-0.3cm}
			\caption*{by Spacial Distribution of Air Pollution Measurement Source}
			\label{main_space}
			\begin{tabular}{lccccccccccccc}
				\toprule \toprule
				& \multirow{2}{*}{Localized} & & \multicolumn{3}{c}{Distant Measure} \\
				\cline{4-6}
				& \multirow{2}{*}{Measure} & & The Closest & Two Closest & Three Closest \\
				& & & Station & Stations & Stations \\
				\cline{2-6}
				&  (1) & & (2) & (3) & (4) \\
				\midrule
			
	$PM_{10}$  ($\mu g/m^{3}$)  &      -0.011$^{***}$ &   &    0.002   &       0.002   &       0.002 \\
	\vspace{0.4cm}
	\hspace{3mm}  &     (0.003)  &  &   (0.002)   &       (0.003)    &     (0.003)   \\

	AQI &      -0.012$^{***}$ &   &    0.003 &       0.002 &       0.002\\
	\vspace{0.4cm} 
	\hspace{3mm} &     (0.004)  &  &   (0.002) &     (0.003) &     (0.003)\\
				
	High pollution ($PM_{10}$) &     -0.181$^{***}$ &  &    -0.010 &      -0.002 &      0.000\\
	\vspace{0.4cm} 
	\hspace{3mm}  &   (0.041) &   &  (0.025)  &     (0.025)  &     (0.025) \\
	
	AQI mean &        8.70   &   &     19.99   &        19.95   &        19.99   \\	
	\vspace{0.2cm}
	Average distance (km) &       -   &    &    16.05   &        21.11   &        25.64   \\			
	Number of observations &        4,650   &    &     4,650   &         4,650   &         4,650  \\
	\bottomrule
		\end{tabular}
			\begin{minipage}{15cm}
				\vspace{0.2cm}
					\footnotesize{
					\textit{Notes}: The table shows the estimated effects of air pollution on the standardized exam results, where the air pollution measures are from different sources. In column 1, air pollution is measured directly in the schoolyard by the devices installed as part of the ESA program. These results are identical to the those presented in column 4 of Table \ref{main_results}. In columns 2--4, air pollution is measured by public air pollution measuring stations. In column 2, measurements from the closest air pollution measuring station to a give school are considered. In columns 3 and 4, measurements from the two and three closest air pollution measuring stations are taken into account. Here, air pollution values are calculated using the inverse distance weighting. At the bottom of the table, the average air pollution (measured in AQI) as well as the average distance (in kilometers) between the schools and the air pollution measuring stations are shown. In the first row, air pollution is measured as the concentration of $PM_{10}$. In the second row, the Air Quality Index (AQI) is calculated based on the $PM_{10}$ values. In the third row, the high pollution dummy is based on $PM_{10}$ values and defined as 1 if the daily air pollution concentration is equal to or greater than 20 $\mu g/m^{3}$ and to 0 otherwise. All regressions include school and time fixed effects, weather controls, nonlinear weather controls (squared weather controls and interaction terms between these weather controls), and school-related control variables. All estimations are based on 4,650 observations and the sample mean of the outcome variable is 0.000. Robust standard errors (presented in parentheses) allow for clustering at the school level. *** p$<$0.01, ** p$<$0.05, * p$<$0.1.}
			\end{minipage}
		\end{center}
	\end{table}


\clearpage

\begin{table}
		\begin{center}
			\caption{Effects of Air Pollution on Exam Results}
			\vspace{-0.3cm}
			\caption*{by Air Pollution Measurement Time Window}
			\label{main_time}
			\begin{tabular}{lccccccccccccc}
				\toprule \toprule
				& 22h Before + & 10h Before + & 2h Before + & \multirow{2}{*}{2h Exam} \\
				& 2h Exam & 2h Exam & 2h Exam \\
				\cline{2-5}
				&  (1) & (2) & (3) & (4) \\
				\midrule
			
	$PM_{2.5}$  ($\mu g/m^{3}$) &      -0.014$^{***}$ &      -0.004 &      -0.001 &      -0.001  \\
	\vspace{0.2cm}
	\hspace{3mm} &     (0.004) &     (0.003) &     (0.003) &     (0.002)\\
	
	$PM_{10}$  ($\mu g/m^{3}$)  &      -0.011$^{***}$ &      -0.004 &      -0.002 &      -0.001 \\
	\vspace{0.2cm}
	\hspace{3mm} &     (0.003) &     (0.002) &     (0.002) &     (0.002)\\

	AQI  &      -0.012$^{***}$ &      -0.004 &      -0.002 &      -0.001\\
	\vspace{0.2cm} 
	\hspace{3mm}  &     (0.004)  &     (0.003)  &     (0.002)  &     (0.002) \\
				
	High pollution ($PM_{2.5}$)  &      -0.206$^{***}$ &      -0.047 &      -0.055$^{*}$ &      -0.062$^{*}$\\
	\vspace{0.2cm} 
	\hspace{3mm} &  (0.037) &     (0.033) &     (0.032) &     (0.033)\\
	
	High pollution ($PM_{10}$) &      -0.181$^{***}$ &      -0.096$^{***}$ &      -0.079$^{**}$ &      -0.058$^{*}$ \\
	\vspace{0.2cm} 
	\hspace{3mm}  &     (0.041) &     (0.033)  &     (0.033)  &     (0.034)\\
	
		\vspace{0.1cm}
		AQI mean &        8.70  &        10.66   &        9.27   &        8.72   \\	
	
		Number of observations &        4,650   &         4,650   &         4,650   &         4,650  \\
	
	\bottomrule
		\end{tabular}
			\begin{minipage}{15cm}
				\vspace{0.2cm}
					\footnotesize{
					\textit{Notes}: The table shows the estimated effects of air pollution on the standardized exam results, where the time window in which the air pollution is measured varies. In column 1, air pollution is calculated as the average air pollution in a 24-hour time window comprising the 2 hours of the examination and 22 hours before it. These results are identical to those presented in column 4 of Table \ref{main_results}. In column 2, air pollution is calculated as the average air pollution in the 12-hour time window comprising the 2 hours of the examination and 10 hours before it. In column 4, air pollution is calculated as the average air pollution in the 4-hour time window comprising the 2 hours of the examination and 2 hours before it. In column 4, air pollution is calculated as the average air pollution in the 2-hour time window comprising the 2 hours of the examination only. At the bottom of the table, the average air pollution (measured in AQI) is shown. In the first and second rows, air pollution is measured as the concentration of $PM_{2.5}$ and $PM_{10}$, respectively. In the third row, the Air Quality Index (AQI) is calculated based on the $PM_{10}$ values. In the fourth row, the high pollution dummy is based on $PM_{2.5}$ values and defined as 1 if the daily air pollution concentration is equal to or greater than 15 $\mu g/m^{3}$, and to 0 otherwise. In the last row, the high pollution dummy is based on $PM_{10}$ values and defined as 1 if the daily air pollution concentration is equal to or greater than 20 $\mu g/m^{3}$, and to 0 otherwise. All regressions include school and time fixed effects, weather controls, nonlinear weather controls (squared weather controls and interaction terms between these weather controls), and school-related control variables. All estimations are based on 4,650 observations and the sample mean of the outcome variable is 0.000. Robust standard errors (presented in parentheses) allow for clustering at the school level. *** p$<$0.01, ** p$<$0.05, * p$<$0.1.}
			\end{minipage}
		\end{center}
	\end{table}

	
	\begin{landscape}
	\end{landscape}
	
	\clearpage
	\begin{appendix}
		\section{Appendix}
		\label{Appendix}
		\numberwithin{table}{section}
		\numberwithin{figure}{section}
		\setcounter{table}{0}
		\setcounter{figure}{0}
		

\clearpage

\begin{figure}[hbt!]
	\begin{center}
		
		\caption{Education System in Poland}
		\label{school_system}
		\includegraphics[width=\linewidth]{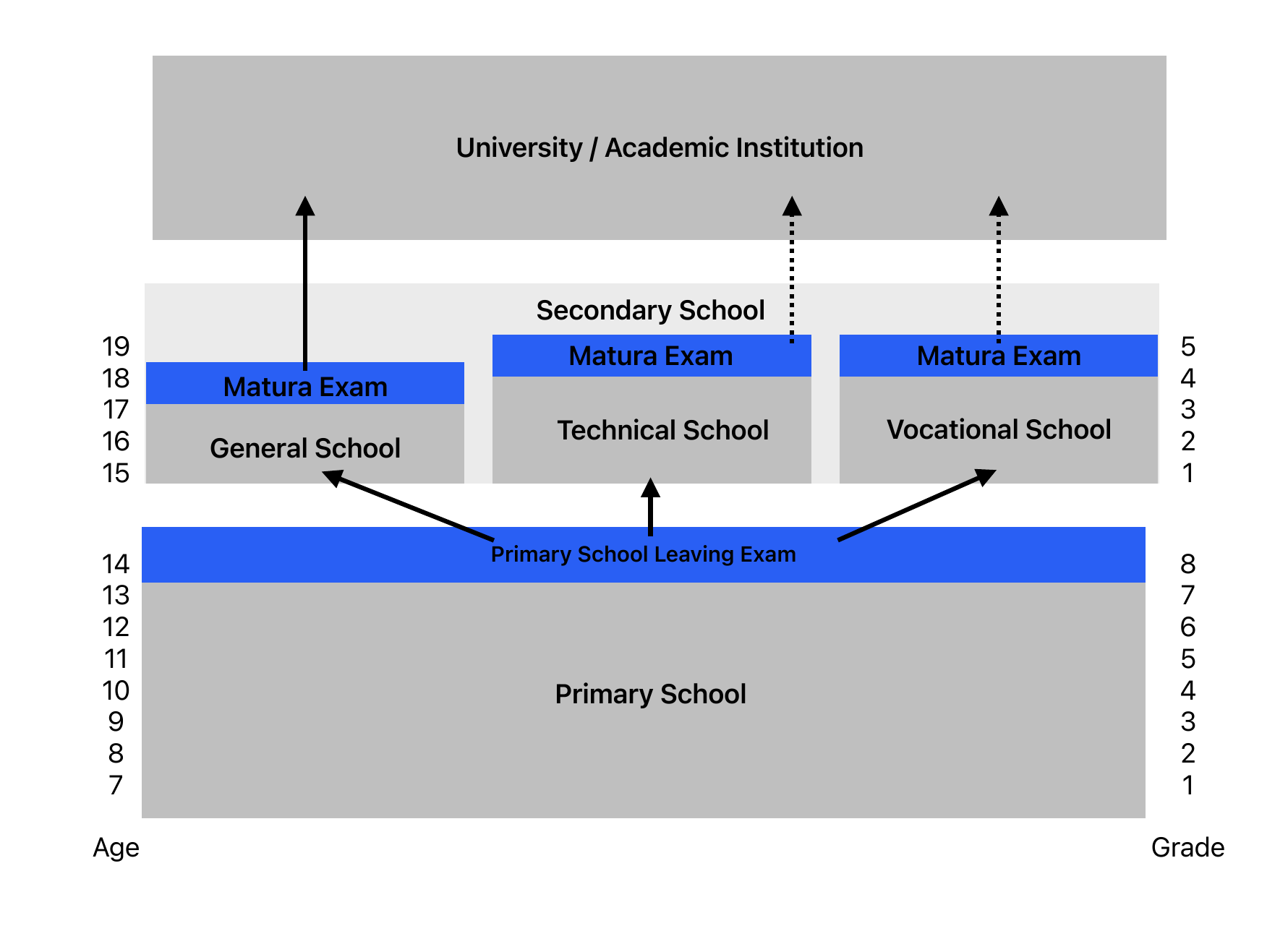}        
		
	\begin{minipage}{12cm}
		\vspace{0.5cm}
		\footnotesize{
			\textit{Notes}: The figure shows the structure of the education system in Poland: mandatory primary school education (children aged 7--14 years), obligatory secondary school education (children aged 15--19 years), and higher education at a university or other academic institution.}
	\end{minipage}
		
\end{center}
\end{figure}


\clearpage

\begin{figure}
\begin{center}
	
		\centering    		
		\caption{Seasonal Pattern in Air Pollution}
		\label{poland_over_year}
		\vspace{0.2cm}
		\centering
	
	\begin{subfigure}{0.48\textwidth}
		\centering
		\caption{Air Pollution Measured in $PM_{2.5}$ (2022)}	
		\includegraphics[width=\textwidth]{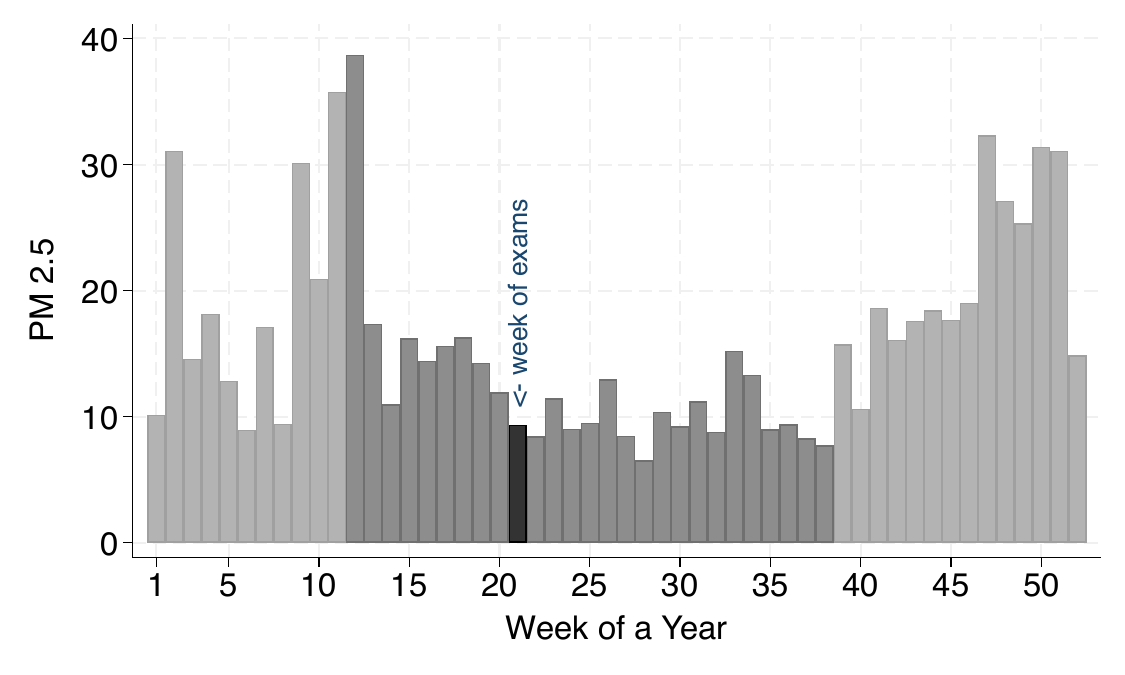}
	\end{subfigure}
	\hspace{0.2cm}
	\begin{subfigure}{0.48\textwidth}
		\centering
		\caption{Air Pollution Measured in $PM_{10}$ (2022)}
		\includegraphics[width=\textwidth]{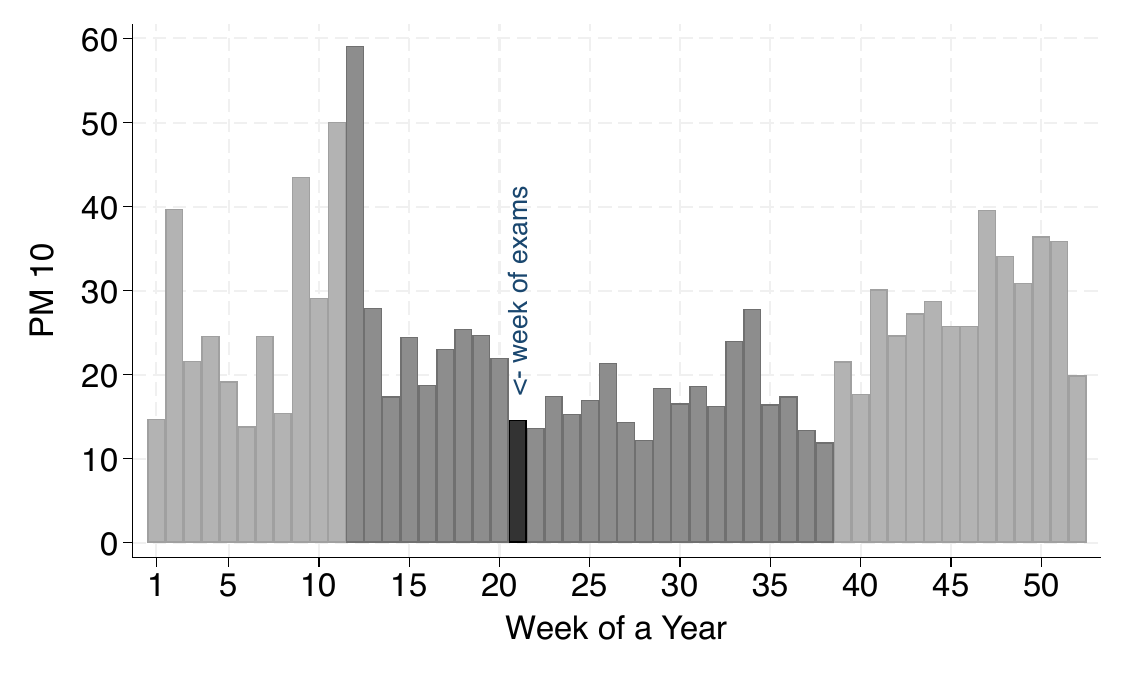}
		\vspace{0.1cm}
	\end{subfigure}

	\begin{subfigure}{0.48\textwidth}
		\centering
		\caption{Air Pollution Measured in $PM_{2.5}$ (2023)}
		\includegraphics[width=\textwidth]{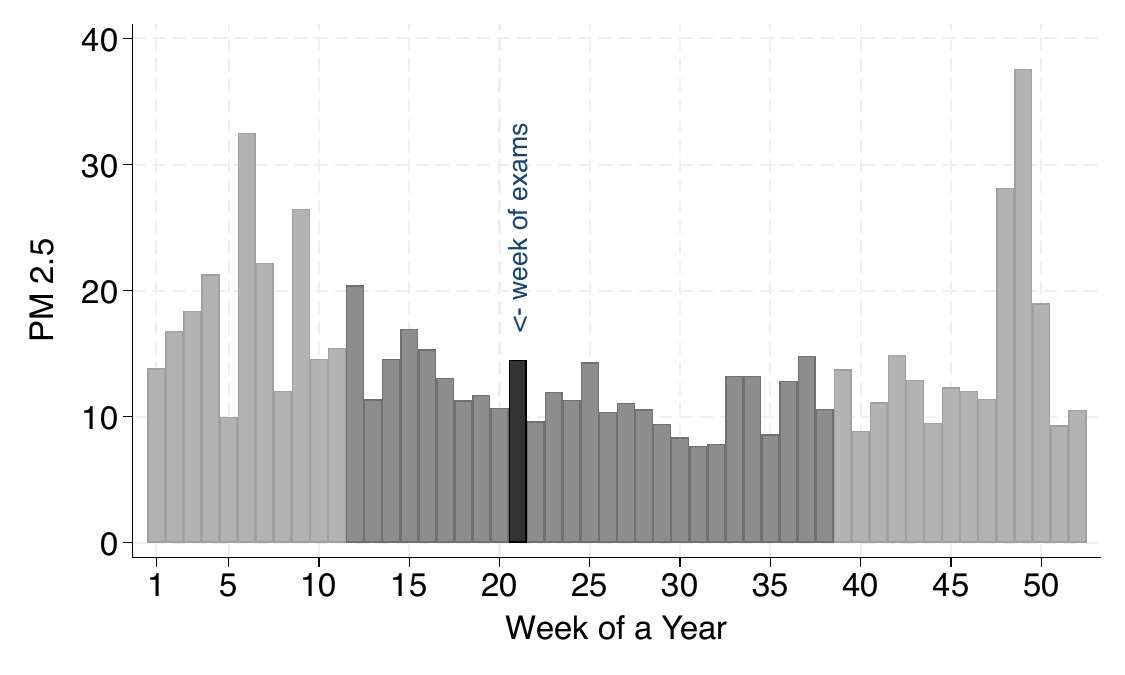}
	\end{subfigure}
	\hspace{0.2cm}
	\begin{subfigure}{0.48\textwidth}
		\centering
		\caption{Air Pollution Measured in $PM_{10}$ (2023)}
		\includegraphics[width=\textwidth]{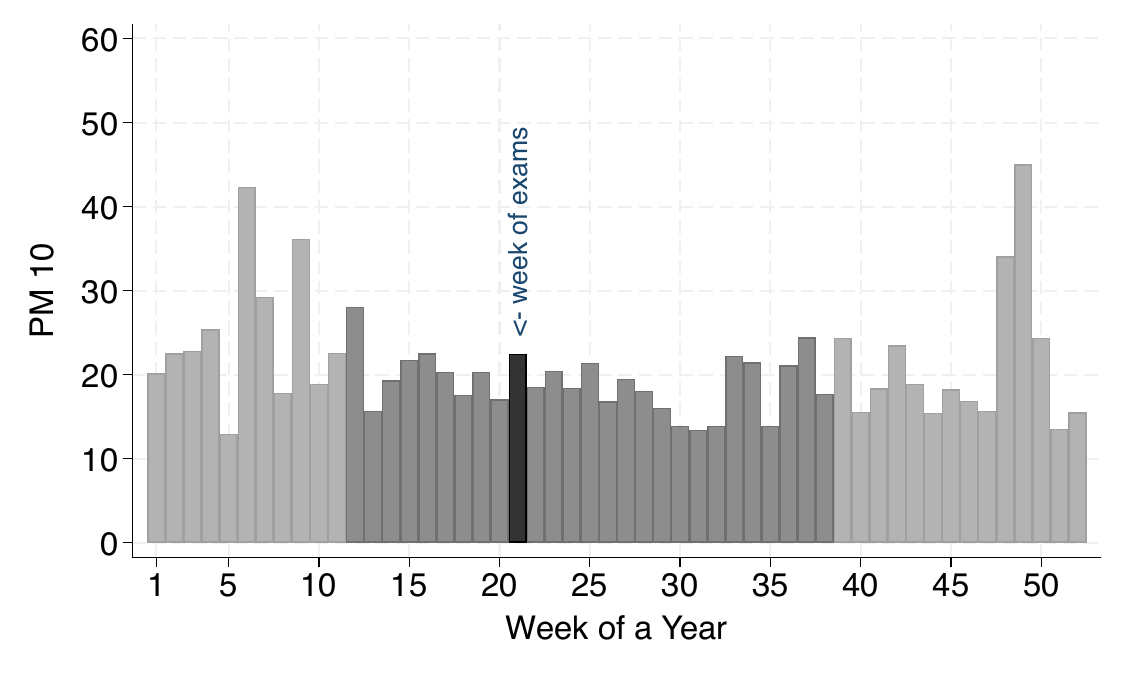}
		\vspace{0.1cm}
	\end{subfigure}
	
	\begin{minipage}{12cm}
		\vspace{0.5cm}
		\footnotesize{
			\textit{Notes}: The figures show the distribution of air pollution measured in $PM_{2.5}$ and $PM_{10}$ throughout the years 2022 and 2023. Air pollution is presented as a national weekly average concentration calculated based on daily measurements taken at all public air pollution measuring stations across the country. The color of the bar indicates the season. Light gray bars refer to the fall/winter season, while dark gray bars refer to the spring/summer season. Black bars indicate the weeks in which the school leaving exams took place. Data on air pollution are not yet available for the whole of 2024, and this year is therefore not shown on the figure.}
	\end{minipage}
	
\end{center}	
\end{figure}


\clearpage

\begin{figure}
	\begin{center}
		
		\caption{Time Series of Exposure to Air Pollution \\ for Randomly Selected Schools}
		\vspace{0.2cm} 
		\label{example}
		
		\begin{subfigure}{0.85\textwidth}
			\caption{Air Pollution Measured in $PM_{2.5}$}
			\includegraphics[width=\linewidth]{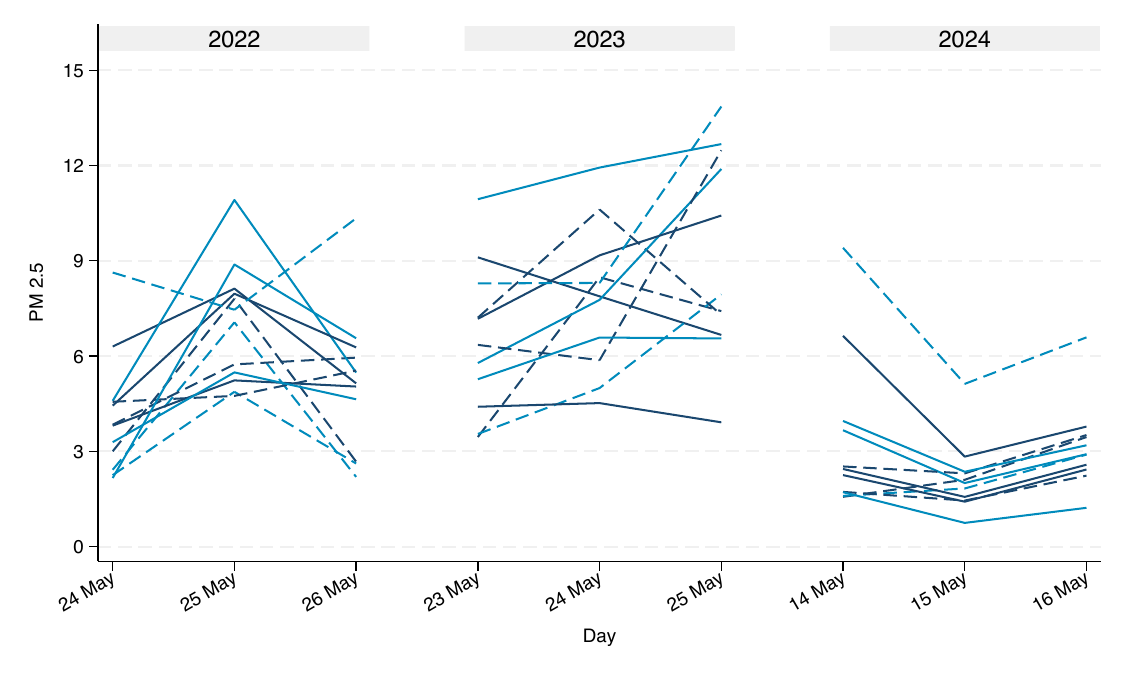}        
		\end{subfigure}
		\hfill		
		
		\begin{subfigure}{0.85\textwidth}
			\caption{Air Pollution Measured in $PM_{10}$}
			\includegraphics[width=\linewidth]{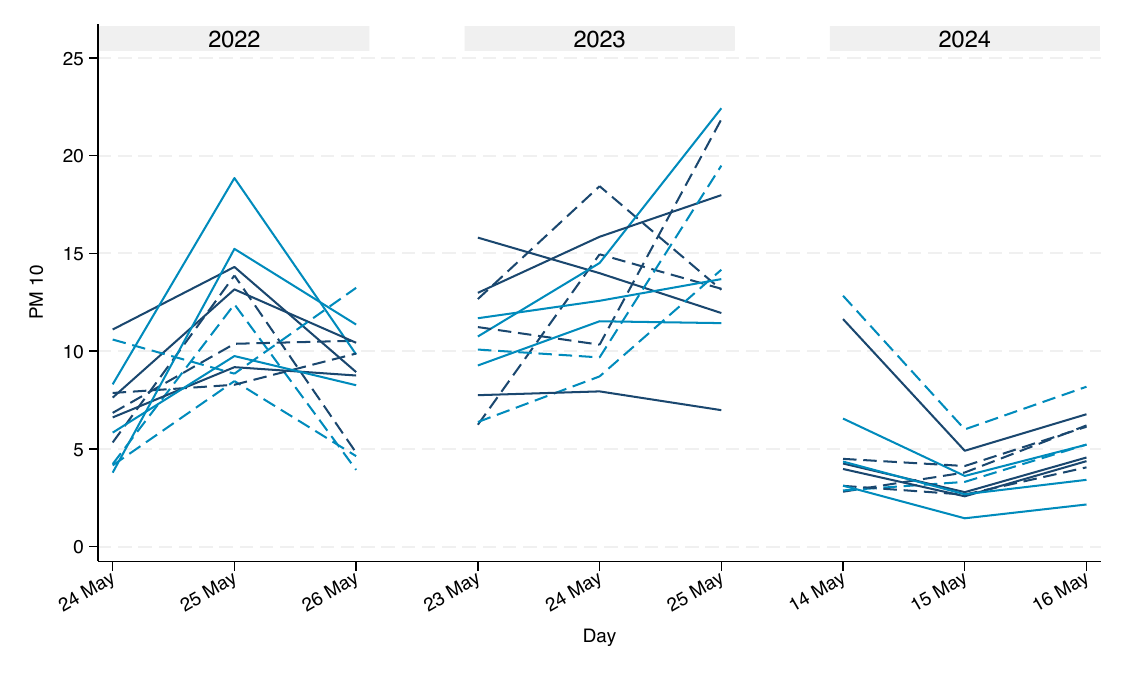}        
		\end{subfigure}
		
	\begin{minipage}{12cm}
		\vspace{0.3cm}
		\footnotesize{
			\textit{Notes}: The figure plots the time series of air pollution across the days of the exams for 12 randomly selected schools within the predefined clusters. The clusters are defined based on the urbanization status of the municipality in which a given school is located (urban vs. rural municipality) and typical air pollution in the province in which a given school is located (polluted vs. cleaner provinces). Within each cluster, three schools, one from each of the three quantiles of the school size distribution, are randomly chosen. In the top figure (a), air pollution is measured in $PM_{2.5}$ and in the bottom figure (b), it is measured in $PM_{10}$.}
	\end{minipage}
		
\end{center}
\end{figure}


\clearpage

\begin{figure}
	\begin{center}
		
		\caption{Plot of Residual Exam Result and Residual Air Pollution}
		\vspace{0.2cm} 
		\label{residuals}
		
		\begin{subfigure}{0.85\textwidth}
			\caption{Air Pollution Measured in $PM_{2.5}$}
			\includegraphics[width=\linewidth]{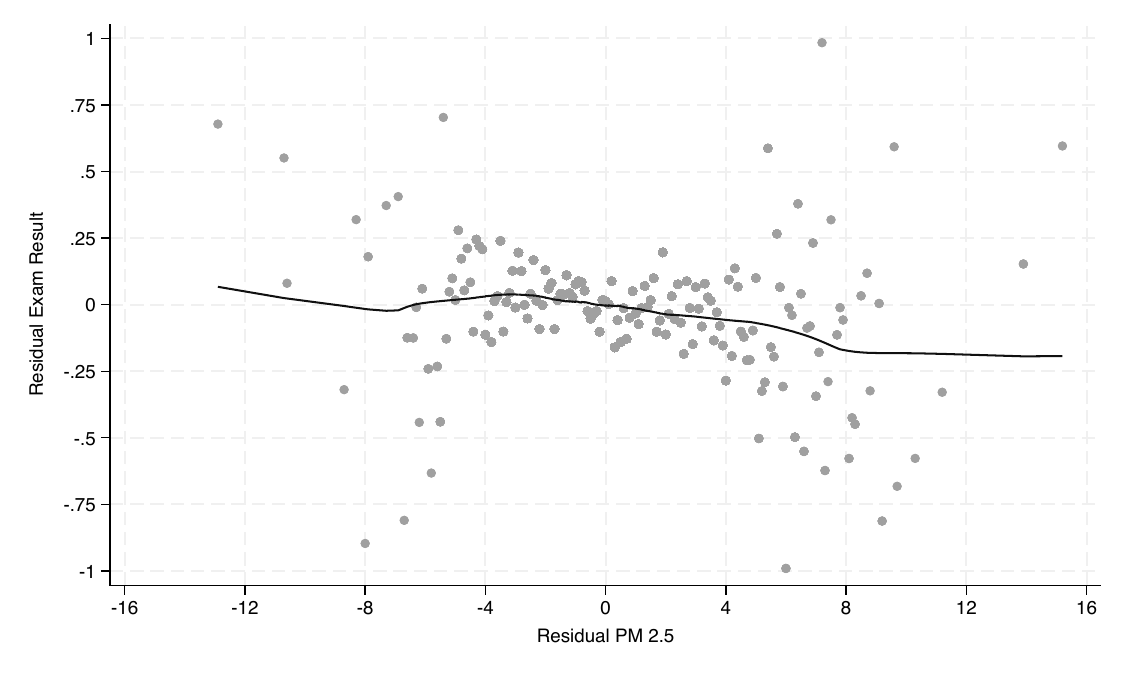}        
		\end{subfigure}
		\hfill		
		
		\begin{subfigure}{0.85\textwidth}
			\caption{Air Pollution Measured in $PM_{10}$}
			\includegraphics[width=\linewidth]{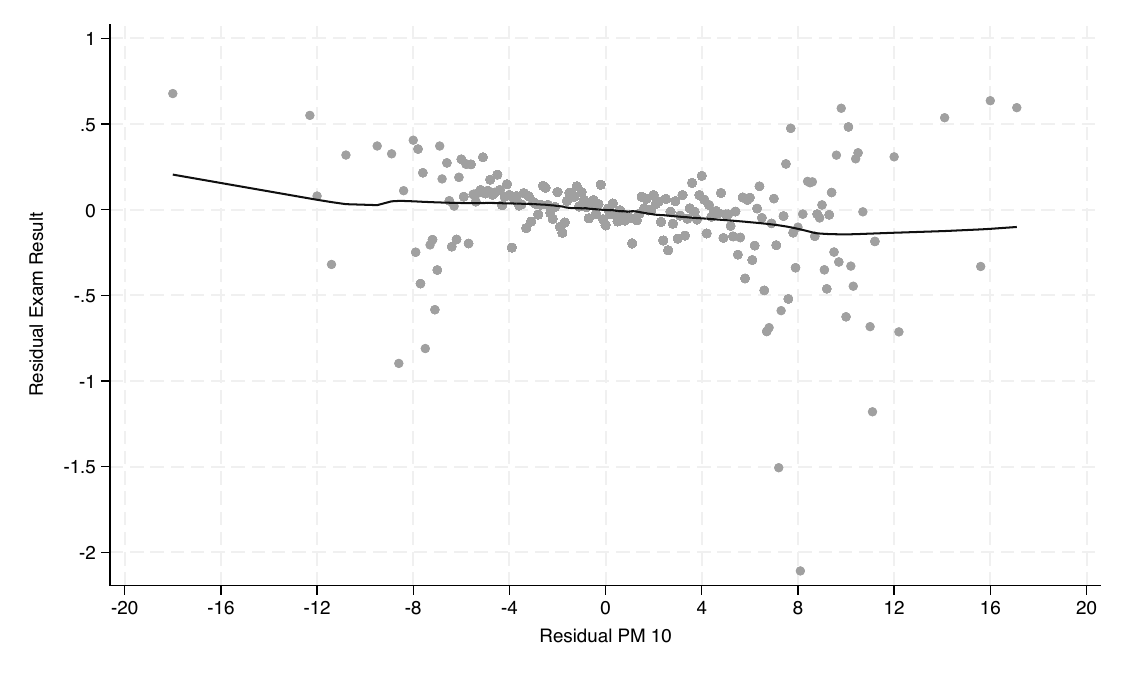}        
		\end{subfigure}
		
	\begin{minipage}{12cm}
		\vspace{0.5cm}
		\footnotesize{
			\textit{Notes}: The figures plot the residuals from the regression of standardized exam results on school fixed effects, time fixed effects, and all other control variables used in the main specification (y axis) against the residuals from the regression of air pollution on the same set of variables (x axis). The standardized exam results residuals are averaged over 0.1 unit bins of air pollution residuals. The graph is produced using lowess bandsmoother. In the top figure (a), air pollution is measured in $PM_{2.5}$ and in the bottom figure (b), it is measured in $PM_{10}$.}
	\end{minipage}
		
\end{center}
\end{figure}


\clearpage

\begin{figure}
	\begin{center}
		
		\caption{Relation Between Air Pollution and Exam Results}
		\vspace{0.2cm} 
		\label{polution_vs_quality}
		
		\begin{subfigure}{0.85\textwidth}
			\caption{Air Pollution Measured in $PM_{2.5}$}
			\includegraphics[width=\linewidth]{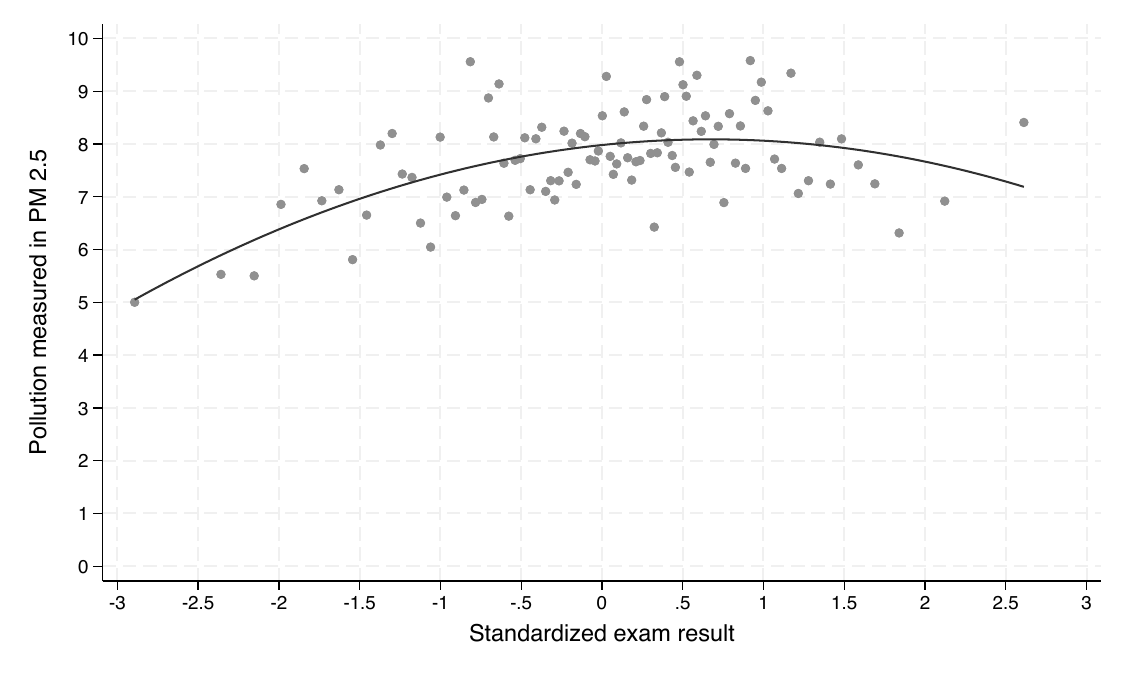}        
		\end{subfigure}
		\hfill		
		
		\begin{subfigure}{0.85\textwidth}
			\caption{Air Pollution Measured in $PM_{10}$}
			\includegraphics[width=\linewidth]{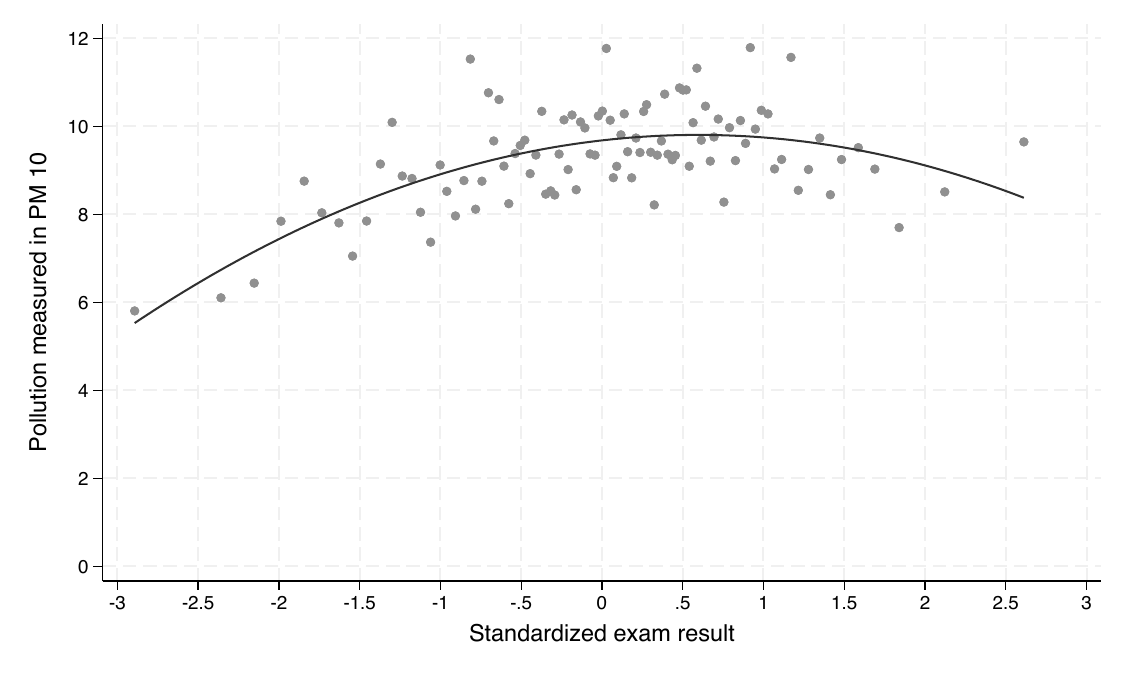}        
		\end{subfigure}
		
	\begin{minipage}{12cm}
		\vspace{0.5cm}
		\footnotesize{
			\textit{Notes}: The figures show the relationship between air pollution on the days of the exams (y axis) and the standardized average exam result of the school in the three obligatory subjects (x axis). The black curves represent the quadratic fit lines. In the top figure (a), air pollution is measured in $PM_{2.5}$, while in the bottom figure (b), it is measured in $PM_{10}$.}
	\end{minipage}
		
\end{center}
\end{figure}


\clearpage

\begin{figure}
	\begin{center}
		
		\caption{Location of Public Air Pollution Measuring Stations}
		\vspace{0.2cm} 
		\label{location_stations}
		
		\begin{subfigure}{0.85\textwidth}
			\caption{$PM_{2.5}$ Measuring Stations}
			\includegraphics[width=\linewidth]{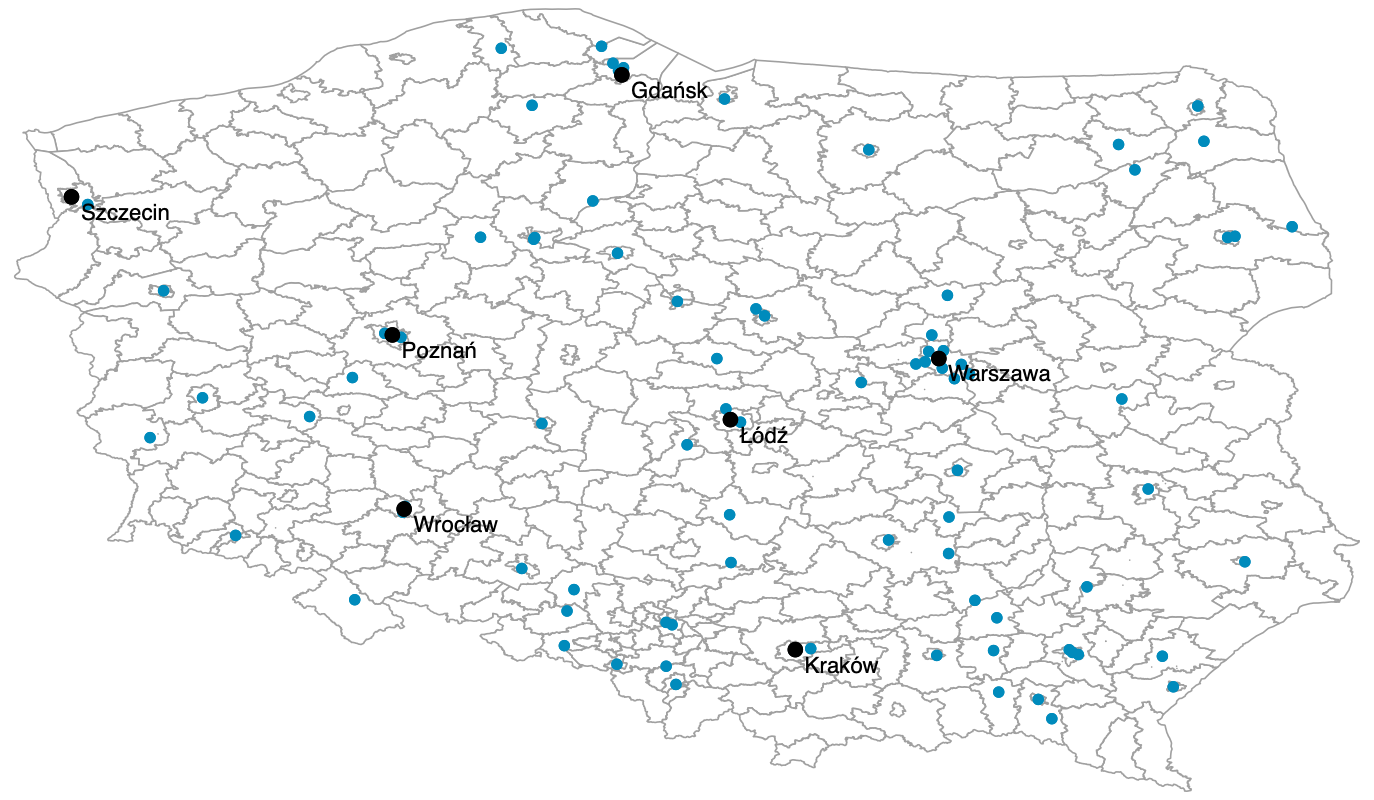}        
		\end{subfigure}
		\hfill		
		
		\begin{subfigure}{0.85\textwidth}
			\caption{$PM_{10}$ Measuring Stations}
			\includegraphics[width=\linewidth]{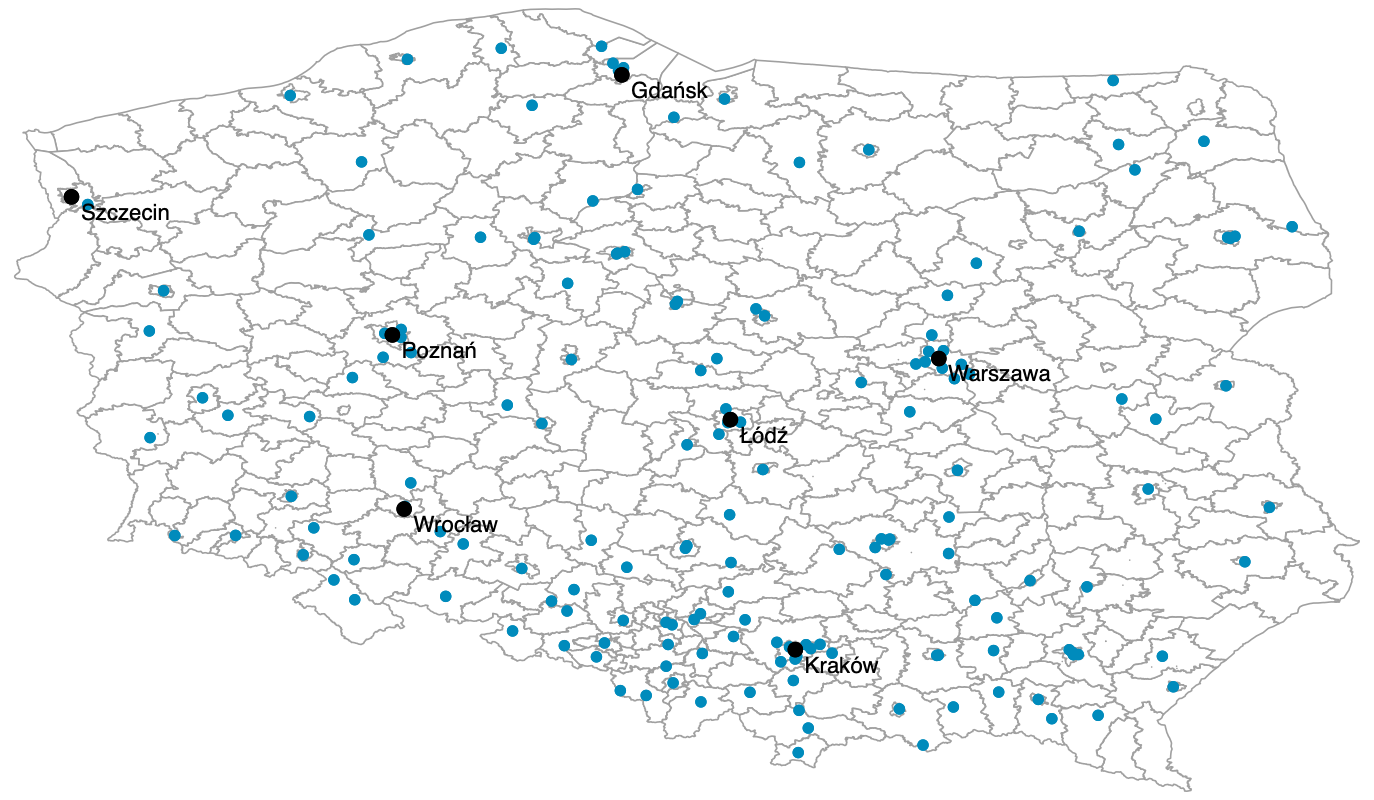}        
		\end{subfigure}
		
	\begin{minipage}{12cm}
		\vspace{0.5cm}
		\footnotesize{
			\textit{Notes}: The upper figure (a) shows the location of the measuring stations where  $PM_{2.5}$ is measured. The bottom figure (b) shows the location of the measuring stations where $PM_{10}$ is measured. Major cities are marked for clarity.}
	\end{minipage}
		
\end{center}
\end{figure}


\clearpage

\begin{figure}
	\begin{center}
		
		\caption{Effects of Air Pollution on Exam Results}
		\vspace{-0.3cm}
		\caption*{Within Specified School--Air Pollution Measuring Station Distances}
		\vspace{0.09cm} 
		\label{effect_by_disatnce}
		
		\begin{subfigure}{0.55\textwidth}
			\caption{Air Pollution Measured in $PM_{10}$}
			\includegraphics[width=\linewidth]{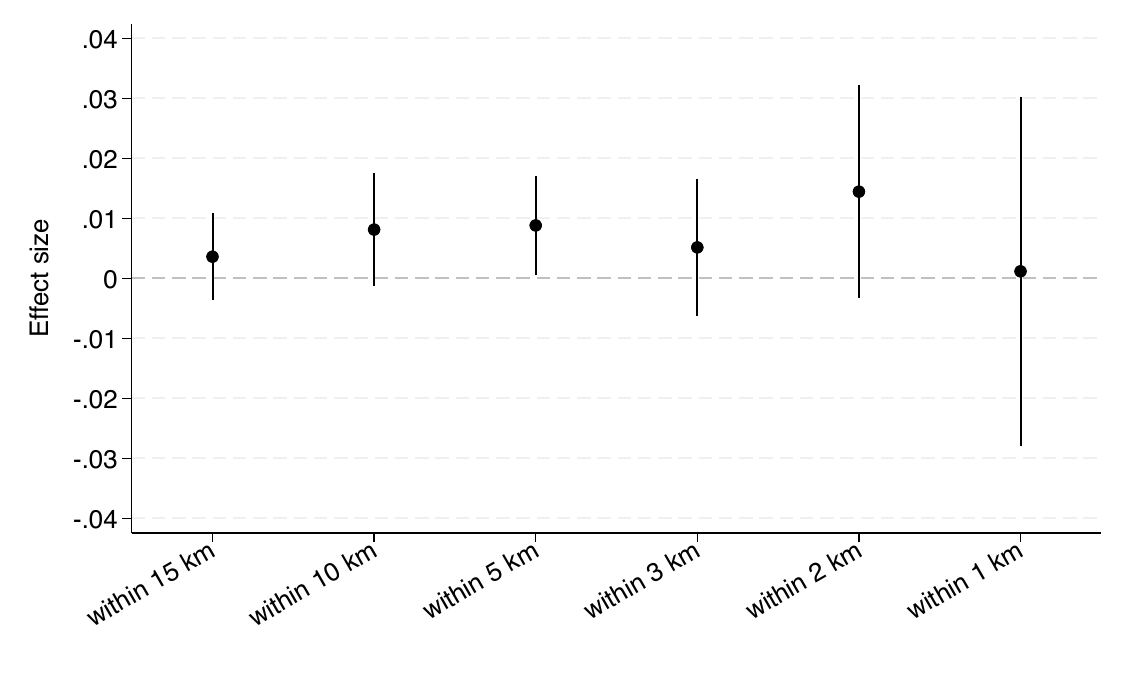}        
		\end{subfigure}
		\hfill		
		
		\begin{subfigure}{0.55\textwidth}
			\caption{Air Pollution Measured as AQI}
			\includegraphics[width=\linewidth]{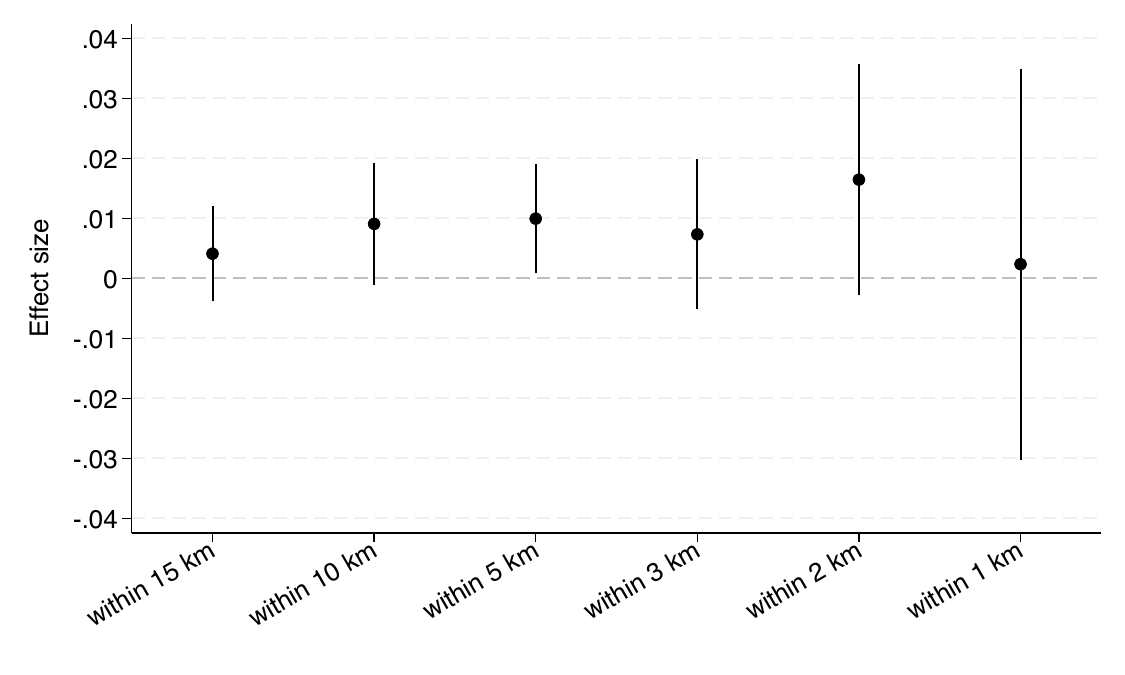}        
		\end{subfigure}
		\hfill
		
		\begin{subfigure}{0.55\textwidth}
			\caption{High Pollution ($PM_{10}$)}
			\includegraphics[width=\linewidth]{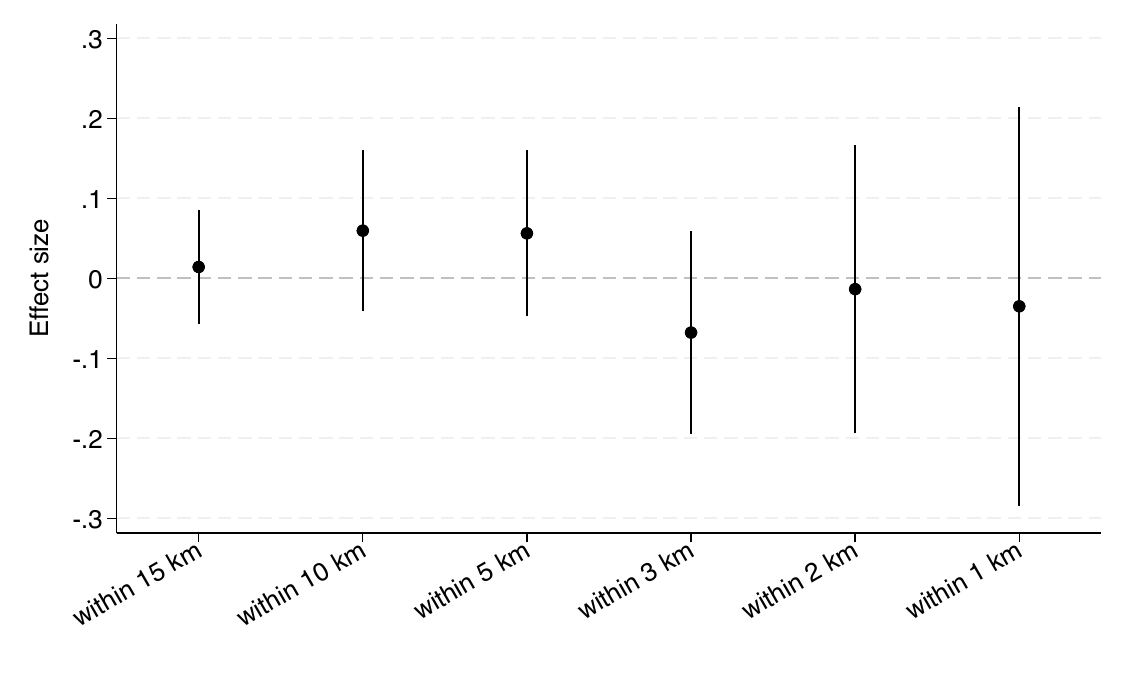}        
		\end{subfigure}
		
	\begin{minipage}{16.5cm}
		\vspace{0.1cm}
		\footnotesize{
			\textit{Notes}: The figures show the effects of air pollution on the standardized exam results together with 95\% confidence intervals for schools located within a specified distance from the closest air pollution measuring station. In the top figure (a), air pollution is measured in $PM_{10}$. In the middle figure (b), air pollution is measured as AQI. In the bottom figure (c), air pollution is measured by a high pollution dummy variable equal to 1 if the daily concentration of $PM_{10}$ is equal to or greater than 20 $\mu g/m^{3}$, and to 0 otherwise. The sample size for the regression with schools located up to 15 km from the closest air pollution measuring station is 2,542, for schools up to 10 km away 1,748, for schools up to 5 km away 1,102, for schools up to 3 km away 707, for schools up to 2 km away 436, and for schools up to 1 km away 117.}
	\end{minipage}
		
\end{center}
\end{figure}


\clearpage

\begin{figure}
\begin{center}
		
		\caption{Air Pollution Measured by ESA Devices \\ and by Public Measuring Stations}
		\label{int_vs_ext}
		\includegraphics[scale=0.6]{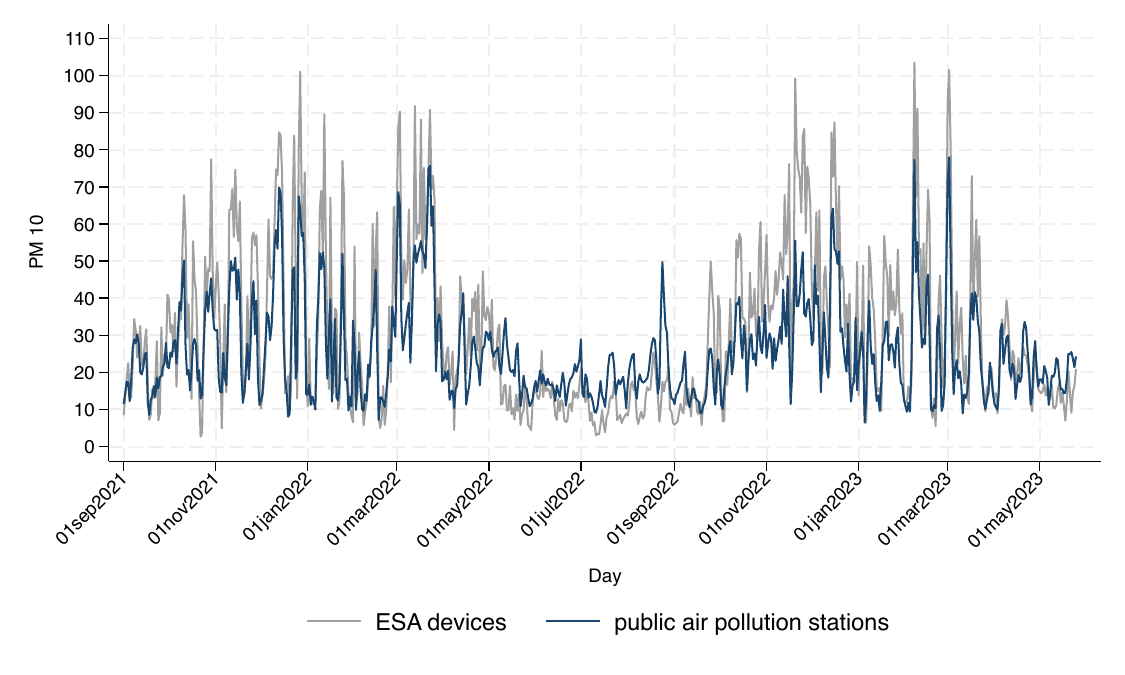}
		
	\begin{minipage}{12cm}
		\vspace{0.5cm}
		\footnotesize{
			\textit{Notes}: The figure shows the relationship between the air pollution ($PM_{10}$) measured by the ESA devices installed in the schoolyard and by the air pollution measuring station closest to a given school. Both air pollution measures are aggregated such that a daily average air pollution for the pool of all considered schools and for the pool of all measuring stations is calculated. Air pollution measured by the air pollution measuring stations is weighted by the inverse of the distance between a measuring station and a school. The gray line represents the air pollution measured by ESA devices, while the blue line measures air pollution measured by the closest measuring stations. The time period considered covers school years 2021/22 and 2022/23. Since the data on air pollution measured by pubic measuring stations are not yet available for the whole school year 2023/24, this is not added to the figure.}
	\end{minipage}
	
\end{center}
\end{figure}


\clearpage

\begin{figure}
	\begin{center}
		
		\caption{Hourly Variation in Air Pollution on the Exam Days}
		\vspace{0.2cm} 
		\label{over_day_time}
		
		\begin{subfigure}{0.85\textwidth}
			\caption{Air Pollution Measured in $PM_{2.5}$}
			\includegraphics[width=\linewidth]{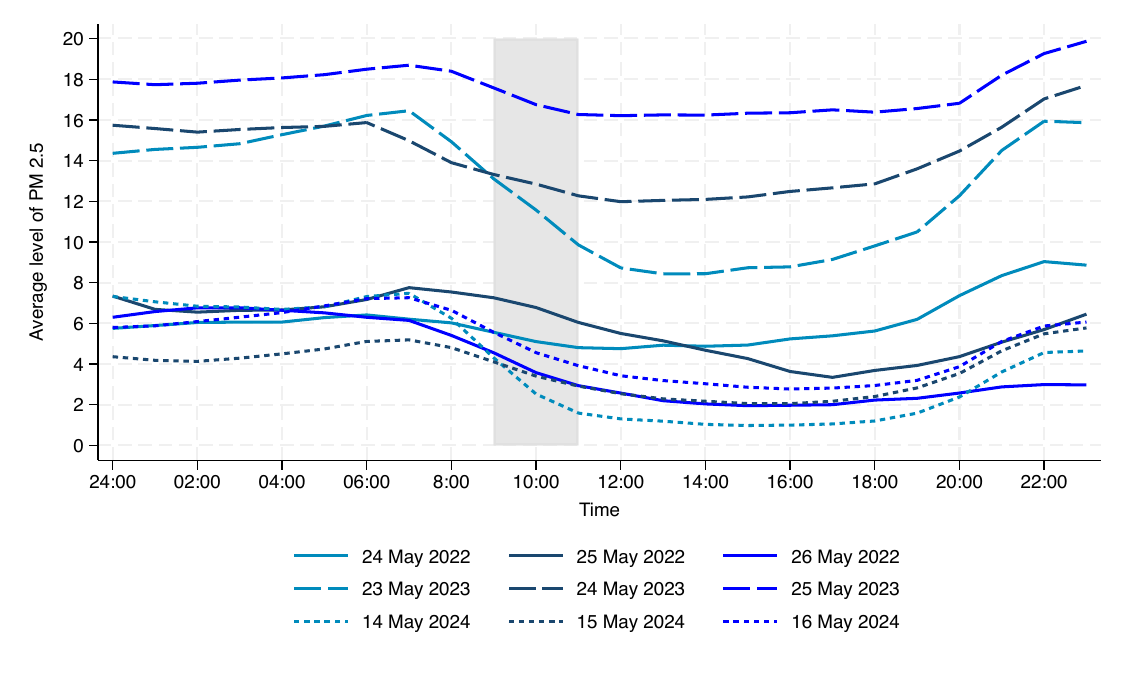}        
		\end{subfigure}
		\hfill		
		
		\begin{subfigure}{0.85\textwidth}
			\caption{Air Pollution Measured in $PM_{10}$}
			\includegraphics[width=\linewidth]{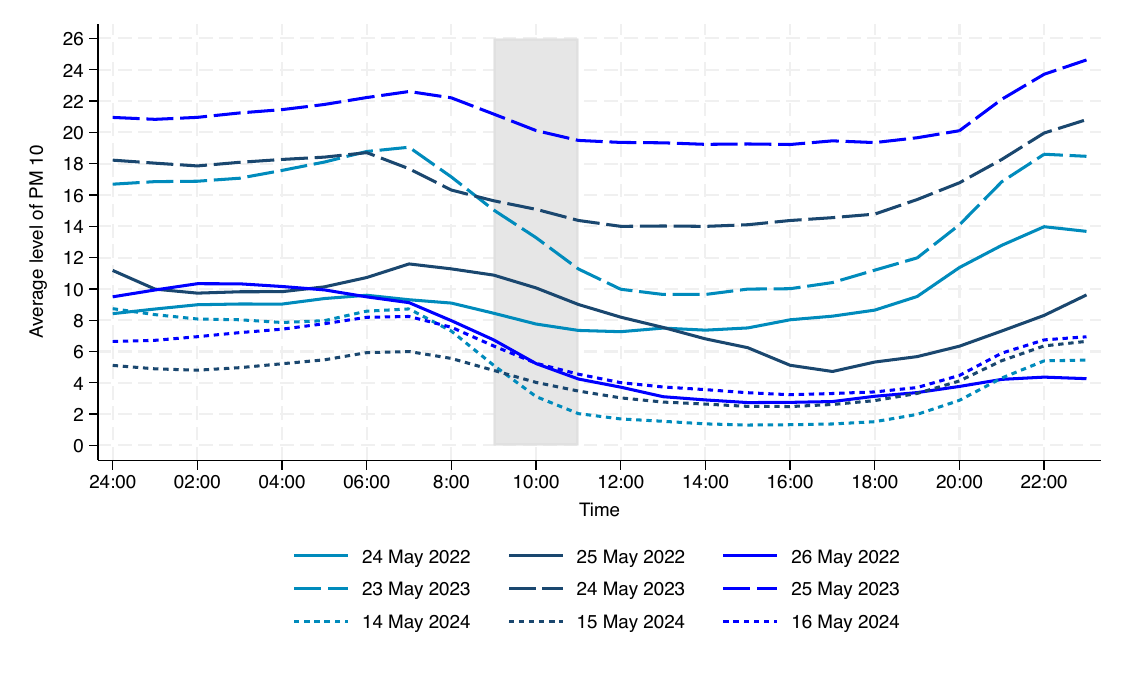}        
		\end{subfigure}
		
	\begin{minipage}{12cm}
		\vspace{0.5cm}
		\footnotesize{
			\textit{Notes}: The figures show the evolution of air pollution over the course of the days on which school leaving exams took place in the sample of schools participating in the ESA project. In the top figure (a), air pollution is measured in $PM_{2.5}$, while in the bottom figure (b), it is measured in $PM_{10}$. The examination time (9:00--11:00 a.m.) is marked in gray.}
	\end{minipage}
		
\end{center}
\end{figure}


\clearpage

\begin{landscape}

\begin{table}
\small
	\begin{center}	
		\caption{Literature Review}
			\label{literature}
				\begin{tabular}{lllllllll}
					\toprule \toprule
	\addlinespace
	Scientific Paper & Setting & Outcome & Air Pollution & Air Pollution Measure & Effect Size \\  
	\midrule
		
	 Carneiro et al. & Rio de Janerio & university entrance exams: & $PM_{10}$:  & inverse distance weighted & 1 SD increase in $PM_{10}$ \\
	 (2021)  & and S\~ao Paulo, & mean: 526 points & mean: 17.5 $\mu g/m^{3}$ & (IDW) daily average from three & decreases students' \\	 
	  & Brazil & std. dev.: 76.2 & std. dev.: 6.11 & state-run pollution measuring & score by 0.049 SD \\	  
	  & & & & stations within 10 km from the & \\	
	  & &  & & exam's municipality centroid &  \\	
	  
	  & \\
	  
	  Ebenstein et al. & Israel & matriculation exams: & $PM_{2.5}$: & daily average from all state-run & 1 SD increase in $PM_{2.5}$ \\
	  (2016) & & mean: 70.76 points & mean: 21.05 $\mu g/m^{3}$ & pollution measuring stations & decreases students' \\
	  & & std. dev.: 23.74 & std. dev.: 10.86 & within 2.5 km of the exam's & score by 0.038 SD \\
	  & & & & city boundary \\
	  
	  & \\
	  
	  Andersen et. al. & Denmark & upper secondary & $PM_{2.5}$: & daily concentration at & 1 SD increase in $PM_{2.5}$ \\
	  (2024) & & school exit exams & mean: 7.9 $\mu g/m^{3}$ & 1 km
x 1 km grid cells & decreases Math test score \\
	  & & mean: 0.0 & std. dev.: 2.8 & corresponding to & by 0.024 SD and reading \\
	  & & std. dev: 1.0 & & students' home addresses & test score by 0.013 SD \\
	  & &  & & measured by high-resolution \\
	  & & & & modeling system \\
 
	  & \\
	  
	  Sefi Roth & London, & university final & $PM_{10}$: & self-collected measure & 1 SD increase in $PM_{10}$\\
	  (2015) & Great Britain & examination: & mean: 33.35 $\mu g/m^{3}$ & of indoor air pollution & decreases students' \\
	  & & mean: 54.59 points & std. dev.: 21.51 & at the time of the exam & score by 0.072 SD  \\
	  & & std. dev.: 18.05 & & & \\
	
	\bottomrule
		\end{tabular}
			\begin{minipage}{21.5cm}
				\vspace{0.2cm}
					\footnotesize{
					\textit{Notes}: The table summarizes the most important features of studies addressing the question of the short-term effects of air pollution on cognitive performance in high-stakes settings.}
			\end{minipage}
		\end{center}
	\end{table}
	
\end{landscape}


\clearpage

\begin{table}
	\begin{center}	
		\caption{Representativeness of ESA Schools}
			\label{representativeness}
								\begin{tabular}{lcccccc}
					\toprule \toprule
	& Sample & Population & p-value \\ 
	\cline{2-4}
	&  (1)       & (2)  & (3) \\

	\midrule
			
	\textit{Panel A: Outcome variable} \\	
	\vspace{0.2cm}
	\hspace{2mm} Pooled standardized exam result&       0.023&       0.000 & 0.038 \\
								
	\vspace{0.1cm}
	\textit{Panel B: School location} \\	
	\vspace{0.05cm}
	\hspace{2mm} Polluted region&       0.524&       0.401 & 0.000\\
	\vspace{0.2cm}
	\hspace{2mm} Urban region &       0.434&       0.400 & 0.308\\
	
	\vspace{0.1cm}			
	\textit{Panel C: School size and composition} \\	
	\vspace{0.05cm}
	\hspace{2mm} Number of 8th grade students&      49.577&      40.356 & 0.001 \\
	\vspace{0.2cm}
	\hspace{2mm} Share of females 8th grade students&      49.721&      48.819 & 0.131 \\
				
	\vspace{0.1cm}	
	\textit{Panel D: Remaining characteristics} \\	
	\vspace{0.05cm}
	\hspace{2mm} Public school&       0.962&       0.911 & 0.001\\
	\vspace{0.05cm}
	\hspace{2mm} Independent school &       0.768&       0.767 & 0.885 \\
	\vspace{0.05cm}
	\hspace{2mm} Special school &       0.003&       0.013 & 0.000 \\
	\vspace{0.05cm}
	\hspace{2mm} Special needs school &       0.003&       0.012 & 0.204 \\
	\vspace{0.2cm}
	\hspace{2mm} Boarding school &       0.008&       0.012 & 0.396\\
		
	Number of observations &         788&        12,409\\
	
	\bottomrule
		\end{tabular}
			\begin{minipage}{13cm}
				\vspace{0.2cm}
					\footnotesize{
					\textit{Notes}: The table compares the main characteristics of the schools included in the sample (column 1) to the main characteristics of all primary schools in the country (column 2). Column 3 shows the p-value from the regression of a sample inclusion dummy on each of these characteristics. Panel A shows the sample average pooled standardized exam result. The pooled standardized exam result has a mean of zero and a standard deviation of one in the sample of all schools in the country. Panel B compares the schools' locations. Polluted region is a dummy variable equal to 1 if a school is located in a province considered typically more polluted, and to 0 otherwise. Urban region is a dummy variable equal to 1 if a school is located in an urban municipality, and to 0 otherwise. Panel C relates to the school's size and composition. Panel D refers to the remaining school characteristics. Public school is a dummy variable equal to 1 if a school is public, and to 0 otherwise. Independent school is a dummy variable equal to 1 if a school is not part of a larger educational establishment, and to 0 otherwise. Special school is a dummy variable equal to 1 if a school is of a special type (e.g., aimed at socially maladjusted students), and to 0 otherwise. Special needs school is a dummy variable equal to 1 if a school is aimed at students with physical/intellectual disabilities, and to 0 otherwise. Boarding school is a dummy variable equal to 1 if a school takes boarders, and to 0 otherwise. All school characteristics are measured in school year 2021/22. The number of schools in the sample is 788 and in the whole country 12,409.}
			\end{minipage}
		\end{center}
	\end{table}
	

\clearpage

\begin{landscape}

\begin{table}
		\begin{center}
			\caption{Quantile Regression: \\ Effects of Air Pollution on Exam Results}
			\label{quantile_table}
			\begin{tabular}{lccccccccc}
				\toprule \toprule
				& \multicolumn{9}{c}{Quantile} \\
				\cline{2-10}
				& 1st & 2nd & 3rd & 4th & 5th & 6th & 7th & 8th & 9th \\
				\cline{2-10}
				&  (1) & (2) & (3) & (4) & (5) & (6) & (7) & (8) & (9) \\
				\midrule
				
	$PM_{2.5}$  ($\mu g/m^{3}$) &      -0.012$^{*}$  &      -0.012$^{**}$ &      -0.013$^{***}$&      -0.013$^{***}$&      -0.014$^{***}$&      -0.014$^{***}$&      -0.015$^{***}$&      -0.015$^{***}$&      -0.016$^{**}$ \\
		\vspace{0.3cm}
		\hspace{3mm}  &     (0.006)   &     (0.005)   &     (0.004)   &     (0.004)   &     (0.003)   &     (0.004)   &     (0.004)   &     (0.005)   &     (0.006)   \\
			
	$PM_{10}$  ($\mu g/m^{3}$) &      -0.011$^{*}$  &      -0.011$^{**}$ &      -0.011$^{***}$&      -0.011$^{***}$&      -0.011$^{***}$&      -0.011$^{***}$&      -0.011$^{***}$&      -0.011$^{***}$&      -0.012$^{**}$ \\
		\vspace{0.3cm}
		\hspace{3mm} &     (0.006)   &     (0.004)   &     (0.004)   &     (0.003)   &     (0.003)   &     (0.003)   &     (0.003)   &     (0.004)   &     (0.005)   \\
				

		No. Observations &  4,650 &  4,650 &  4,650 &  4,650 &  4,650 &  4,650 &  4,650 &  4,650 &  4,650 \\
		
	\bottomrule
		\end{tabular}
			\begin{minipage}{21cm}
				\vspace{0.2cm}
					\footnotesize{
					\textit{Notes}: The table shows the quantile regression of the standardized exam results on air pollution. In columns 1--9, the respective decile of the outcome distribution is used. All regressions include school and time fixed effects, weather controls (temperature, pressure, humidity), nonlinear weather controls (squared weather controls and interaction terms between these weather controls), and school-related control variables. Robust standard errors (presented in parentheses) allow for clustering at the school level. *** p$<$0.01, ** p$<$0.05, * p$<$0.1.}
			\end{minipage}
		\end{center}
	\end{table}
	
\end{landscape}
	

\clearpage

\begin{table}
		\begin{center}
			\caption{Characteristics of Air Pollution Measuring Stations}
			\label{stations_charact}
			\begin{tabular}{lccccccccccccc}
				\toprule \toprule
				& \multicolumn{2}{c}{Air Pollution Measured} \\
				\cline{2-3}
				& $PM_{2.5}$ & $PM_{10}$  \\
				\cline{2-3}
				&  (1) & (2) \\
				\midrule
	
	\vspace{0.1cm}			
	\textit{Panel A: Station type} \\	
	\vspace{0.05cm}
	\hspace{2mm} Background &           0.826&       0.881\\
	\vspace{0.05cm}
	\hspace{2mm} Industry &          0.023&       0.030\\
	\vspace{0.2cm}
	\hspace{2mm} Traffic  &       0.151&       0.089\\					
	\vspace{0.1cm}
	\textit{Panel B: Station location} \\	
	\vspace{0.05cm}
	\hspace{2mm} Urban area  &       0.884&       0.875\\
	\vspace{0.05cm}
	\hspace{2mm} Sub-urban area  &       0.093&       0.071\\
	\vspace{0.2cm}
	\hspace{2mm} Outside urban area  &       0.023&       0.054\\
		
	Number of observations &          86&         168\\
		
	\bottomrule
		\end{tabular}
			\begin{minipage}{10cm}
				\vspace{0.2cm}
					\footnotesize{
					\textit{Notes}: The table shows the main characteristics of air pollution measuring stations subdivided into those where $PM_{2.5}$ is measured (column 1) and those where $PM_{10}$ is measured (column 2). The numbers in the table represent the shares of measuring stations of a particular type and in a particular location.}
			\end{minipage}
		\end{center}
	\end{table}
	
	
\end{appendix}
\end{document}